\documentclass[aps,prl,a4paper,preprintnumbers,tightenlines,superscriptaddress,
twocolumn,floatfix,showpacs,final,nofootinbib,11pt]{revtex4-1}
\usepackage{graphicx}
\usepackage{longtable}
\usepackage{amsmath}
\usepackage{amssymb}
\usepackage{subfigure}
\usepackage{hyperref}
\usepackage{dcolumn}
\usepackage{bm}
\usepackage{color}
\usepackage{cleveref}

\newcommand{\mpl}{{M_{\rm {pl}}}}

\newcommand{\dd}{\, {\rm d}}
\newcommand{\gsim}{\;\mbox{\raisebox{-0.5ex}{$\stackrel{>}{\scriptstyle{\sim}}$}
}\;}
\newcommand{\lsim}{\;\mbox{\raisebox{-0.5ex}{$\stackrel{<}{\scriptstyle{\sim}}$}
}\;}
\newcommand{\ctlimplus}{10^{15}}
\newcommand{\ctlim}{10^{-15}}

\newcommand{\ctlimtt}{6\times10^{-15}}

\allowdisplaybreaks
\usepackage{enumitem}
\setenumerate[1]{label=(\arabic*)}

\begin{document}
\title{Implications of the Neutron Star Merger GW170817 for Cosmological Scalar-Tensor Theories} 
\author{Jeremy Sakstein}
\email[Email: ]{sakstein@physics.upenn.edu}
\affiliation{Department of Physics and Astronomy, Center for Particle Cosmology,
University of Pennsylvania, 209 S. 33rd St., Philadelphia, PA 19104, USA}
\author{Bhuvnesh Jain}
\email[Email: ]{bjain@physics.upenn.edu }
\affiliation{Department of Physics and Astronomy, Center for Particle Cosmology,
University of Pennsylvania, 209 S. 33rd St., Philadelphia, PA 19104, USA}

\begin{abstract}
The LIGO/VIRGO collaboration has recently announced the detection of gravitational waves from a neutron star-neutron star merger (GW170817) and the simultaneous measurement of an optical counterpart (the gamma-ray burst GRB 170817A). The close arrival time of the gravitational and electromagnetic waves limits the difference in speed of photons and gravitons to be less than about one part in $\ctlimplus$. This has three important implications for cosmological scalar-tensor gravity theories that are often touted as dark energy candidates and alternatives to $\Lambda$CDM. First, for the most general scalar-tensor theories---beyond Horndeski models---three of the five parameters appearing in the effective theory of dark energy can now be severely constrained on astrophysical scales; we present the  results of combining the new gravity wave results with galaxy cluster observations. Second, the  combination with the lack of strong equivalence principle violations exhibited by the supermassive black hole in M87, constrains the quartic galileon model to be cosmologically irrelevant.  Finally, we derive a new bound on the disformal coupling to photons that implies that such couplings are irrelevant for the cosmic evolution of the field.
\end{abstract}
\maketitle


The terms \emph{dark energy} and \emph{modified gravity} are closely connected at the most general level; all but the simplest alternatives to the $\Lambda$CDM model typically invoke some modification of general relativity (GR) (see \cite{Clifton:2011jh,Joyce:2014kja,Lombriser:2014dua,Bull:2015stt,Koyama:2015vza} for reviews). The most widely studied of these are scalar-tensor theories where a new scalar $\phi$ mediates an additional gravitational interaction between matter that is suppressed in the solar system by screening mechanisms (see \cite{deRham:2016nuf,Burrage:2016bwy,Burrage:2017qrf,Sakstein:2017pqi} for reviews) but that becomes relevant on cosmological scales. This has motivated an intense theoretical effort towards finding the most general scalar-tensor theory that is pathology free, and the modern approach to dark energy model building can be epitomized by the class of models called \emph{beyond Horndeski} (BH) \cite{Gleyzes:2014dya,Gleyzes:2014qga}. BH theories are a complete and general framework for constructing dark energy/modified gravity models (including commonly studied paragons for modified gravity such as chameleons \cite{Khoury:2003rn} and galileons \cite{Nicolis:2008in}), many of which can accelerate without a cosmological constant (self-accelerate). They are therefore viewed as alternatives to the $\Lambda$CDM cosmological model and there is much effort focused on how well upcoming cosmological surveys will constrain them \cite{Alonso:2016suf}.

BH theories make a striking prediction: the speed of gravitational waves in the cosmological background differs in general from the speed of light \cite{DeFelice:2011bh,Bellini:2014fua,Lombriser:2015sxa,Bettoni:2016mij}. Recently, the LIGO/VIRGO consortium has announced the observation of neutron star merger GW170817 \cite{TheLIGOScientific:2017qsa}, a neutron star-neutron star merger that has been localized to the galaxy NGC 4993, about 40 Mpc from the Milky Way. The simultaneous observation of an optical counterpart (the gamma-ray burst GRB 170817A) by the Fermi gamma-ray telescope \cite{Monitor:2017mdv} and several optical telescopes \cite{GBM:2017lvd} implies that the two speeds can differ by at most one part in $\ctlimplus$, more specifically\footnote{ Note that the sign is positive, i.e. the observation of the gravitational waves before the optical counterpart constrains gravitons to move at a faster speed than photons ($c_T>c$). This therefore probes theories where this effect is predicted, unlike previous bounds that constrain the difference in propagation speeds if $c_T<c$ \cite{Moore:2001bv,Hohensee:2009zk}. } $(c_T^2-c^2)/c^2\le \ctlimtt$, where $c_T$ is the speed of gravitational waves and $c$ is the speed of light (this limit comes from the time lag between the LIGO and Fermi detections). This has severe implications for cosmological scalar-tensor theories that we delineate in this letter. 

Cosmologically, deviations for $\Lambda$CDM that fall into the BH class of models can be parameterized by five free functions of time $\{\alpha_M,\,\alpha_K,\,\alpha_B,\,\alpha_H,\,\alpha_T\}$ \cite{Bellini:2014fua,Gleyzes:2014rba}. These are typically referred to as \emph{the effective theory of dark energy} \cite{Creminelli:2008wc,Gubitosi:2012hu,Bloomfield:2012ff}, and constraining both their values and their cosmological time-dependence is one of the goals of upcoming dark energy missions such as DESI, LSST, Euclid and WFIRST (see \cite{Alonso:2016suf} for example). The first describes the running of the Planck mass and the second the kinetic term for the scalar; we will not discuss these here. The third, $\alpha_B$, describes the kinetic-mixing of the scalar and graviton and the fourth, $\alpha_H$ describes the so-called \emph{disformal} properties of the theory \cite{Bekenstein:1992pj,Sakstein:2014isa,Sakstein:2014aca,Ip:2015qsa,Sakstein:2015jca}. The fifth, $\alpha_T=(c_T^2-c^2)/c^2$ is none other than the fractional difference between the speed of gravitons and photons. 
The observation of optical counterparts therefore implies that this is now known: $\alpha_T\approx0$. 

\begin{figure}
\includegraphics[width=0.45\textwidth]{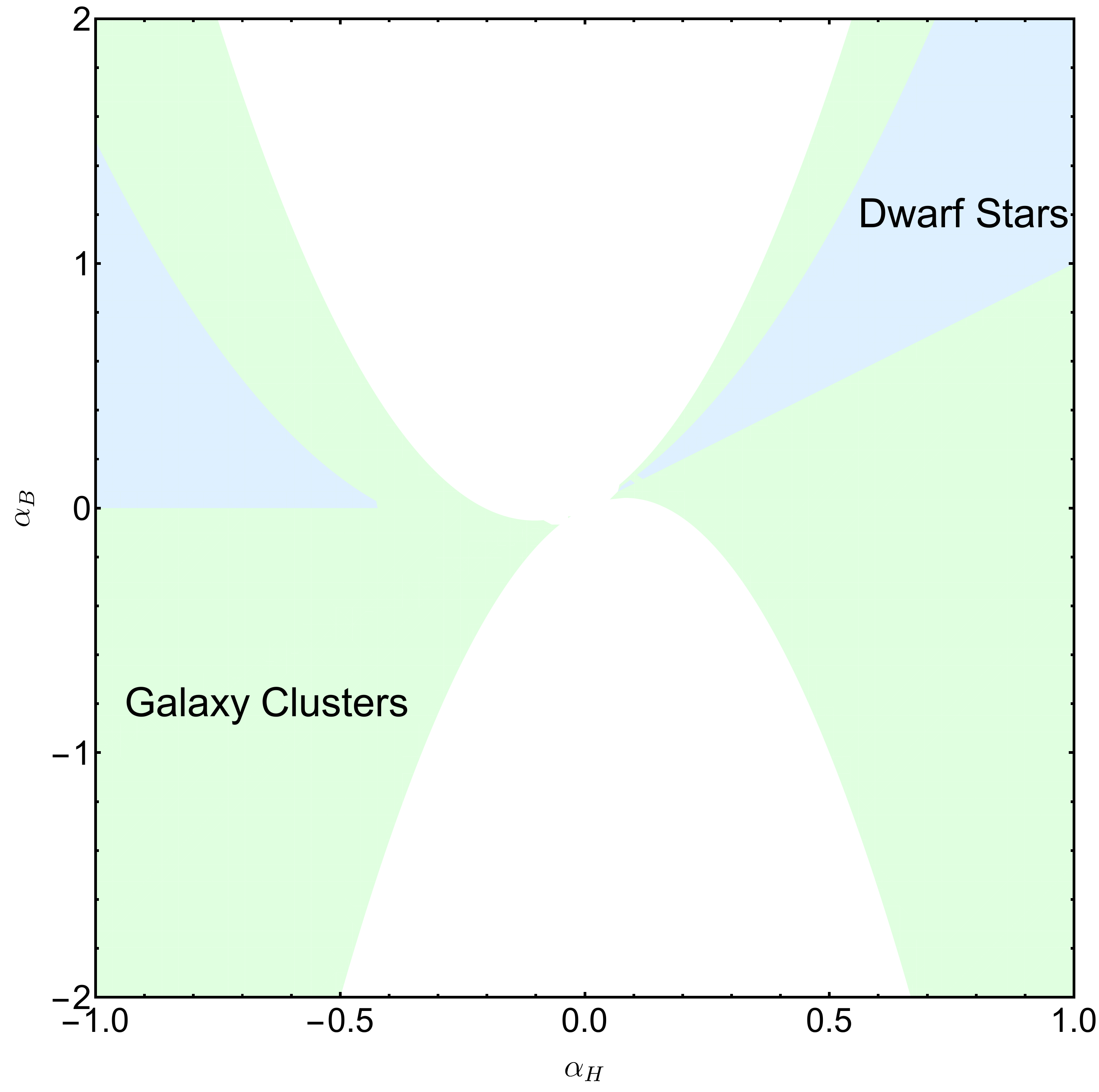}
\caption{The excluded regions in the $\alpha_H$--$\alpha_B$ plane now that $c_T$ is known to be unity with very high precision. The regions excluded by cluster tests and dwarf stars are labeled accordingly. }\label{fig:BH}
\end{figure}

An interesting property of BH theories we will consider\footnote{We are considering quartic theories with $\alpha_H\ne0$ (see \cite{Crisostomi:2017lbg,Langlois:2017dyl,Dima:2017pwp} for discussions of extensions of these). This means we exclude theories that screen using the chameleon mechanism (and similar mechanisms).} is that they satisfy solar system tests of gravity perfectly using the Vainshtein screening mechanism\footnote{Note that the Vainshtein mechanism does not screen deviations in the speed of photons and gravitons \cite{Jimenez:2015bwa}.}, but they predict new and novel deviations from GR inside astrophysical bodies of the form \cite{Kobayashi:2014ida,Koyama:2015oma,Saito:2015fza}:
\begin{align}
\frac{\dd\Phi}{\dd r}&=-\frac{GM(r)}{r^2}-\frac{\Upsilon_1G}{4}\frac{\dd^2 M(r)}{\dd r^2}\\
\frac{\dd\Psi}{\dd r}&=-\frac{GM(r)}{r^2}+\frac{5\Upsilon_2G}{4r}\frac{\dd M(r)}{\dd r},
\end{align}
where
\begin{align}
\Upsilon_1&=\frac{4\alpha_H^2}{(1+\alpha_T)(1+\alpha_B)-\alpha_H-1}\quad\textrm{and}\label{eq:U1}\\
\Upsilon_2&=\frac{4\alpha_H(\alpha_H-\alpha_B)}{5[(1+\alpha_T)(1+\alpha_B)-\alpha_H-1]},\label{eq:U2}
\end{align}
and $\Phi$ and $\Psi$ are the Newtonian potential and the $ij$-component of the metric. This novel \emph{Vainshtein breaking} led to several suggestions for small-scale tests of $\Upsilon_i$ \cite{Koyama:2015oma,Sakstein:2015zoa,Sakstein:2015aac,Sakstein:2015aqx,Jain:2015edg,Sakstein:2016ggl,Babichev:2016jom,Sakstein:2016oel,Sakstein:2016lyj}, which could be used either as priors for cosmological searches or as consistency checks. Note that Eqns. 1 and 2 contain three unknown functions, so until now there was a degeneracy even allowing for the left hand side to be  constrained observationally (as described below). With  $\alpha_T$  known to be negligible these constraints translate uniquely into bounds on $\alpha_B$ and $\alpha_H$ at late times. We present these in Figure \ref{fig:BH}. 

We have used two sets of constraints to obtain these bounds. The first comes from dwarf stars \cite{Sakstein:2015zoa,Sakstein:2015aac}. There is a minimum mass for the onset of hydrogen burning (MMHB) in stars (see \cite{Burrows:1992fg} for a review); stars lighter than this are brown dwarfs while heavier stars are red dwarfs. When $\Upsilon_1>0$, non-relativistic stars are typically less compact so that their cores are cooler and less dense, and therefore this MMHB is larger than the GR value of $0.08M_\odot$. Demanding that the lightest observed red dwarf is at least as heavy as the MMHB sets the bound $\Upsilon_1<1.6$. The second constraint come from galaxy clusters. The equivalence of $\Phi$ and $\Psi$ in GR implies that the mass measured by weak lensing (lensing mass, sensitive to $\Phi + \Psi$) and X-ray data (the surface brightness measured in X-ray data is a probe of the hydrostatic mass, sensitive to $\Phi$) should agree. In BH, this equivalence is broken and so any deviation (or lack thereof) constrains the parameters $\Upsilon_1$ and $\Upsilon_2$. Reference \cite{Sakstein:2016ggl} has performed such a test using weak lensing data from CFHTLenS and X-ray data from XMM-Newton, including measurement errors and systematic uncertainty due to non-thermal pressure (see \cite{Hoekstra:2015gda,Smith:2015qhs,Applegate:2015kua} for studies on the consistency of X-ray and lensing masses). They obtain the constraints $\Upsilon_1=-0.11^{+0.93}_{-0.67}$ and $\Upsilon_2=-0.22^{+1.22}_{-1.19}$. (There is a stronger bound than the lower bound on $\Upsilon_1$, $\Upsilon_1>-2/3$ that we include in Figure \ref{fig:BH}; one cannot form stable stars if this is violated \cite{Saito:2015fza}.)

Figure \ref{fig:BH} shows that a large region of the $\alpha_B$--$\alpha_H$ plane is ruled out\footnote{Note that the cluster constraints of \cite{Sakstein:2016ggl} apply at $z=0.33$ and that the parameters $\alpha_i$ can, in principle, be time-varying. Models where there is a strong variation from $0$ to $1$ between $z=0.33$ and $z=0$ can evade the cluster bounds but not the dwarf star bounds.}. The allowed region can be further constrained by data on galaxy clusters with forthcoming surveys. {Modified gravity models that can explain dark energy typically predict $\alpha_i\sim\mathcal{O}(1)$ so our results are severely constraining for these models. Stage-IV cosmological surveys could constrain $\alpha_i$ to levels of $\mathcal{O}(10^{-1})$ but these make several assumptions about the evolution of these parameters and the amount of screening \cite{Alonso:2016suf}. Our results are independent of these assumptions and are completely general.}. Note that there are two other parameters $\alpha_M$ and $\alpha_K$ that are completely unconstrained on small scales (although one may be able to constrain $\alpha_M$ using tests of the time-variation of Newton's constant). It is also worth noting that the line $\alpha_H=0$ is completely unconstrained, as it should be given equations \eqref{eq:U1} and \eqref{eq:U2}. This line corresponds to a large subset of models known as \emph{Horndeski theories}\footnote{Historically, Horndeski theories are those which give manifestly second-order equations of motion and are therefore free of Ostrogradski instabilities. Later, these were extended to BH theories, which have higher-order equations but propagate three degrees of freedom and are therefore also ghost-free. See \cite{Zumalacarregui:2013pma,Gleyzes:2014qga,Deffayet:2015qwa,Langlois:2015cwa,Crisostomi:2016tcp} for more details and other extensions of the Horndeski theory. We emphasize that the distinction is purely historical, and that Horndeski theories should be considered a subset of the more general BH class.} \cite{Horndeski:1974wa,Deffayet:2011gz}. These theories can still be constrained using gravitational waves, but one needs a second probe since Vainshtein screening works inside objects for these theories. This probe   comes in the form of strong equivalence principle (SEP) violations, which we describe next. 

The entire class of Horndeski theories is vast, and one typically focuses on specific models that encapsulate the relevant physics in order to provide a concrete realization of their cosmological consequences. The quintessential paradigm is the \emph{covariant quartic galileon} \cite{Deffayet:2009wt} with Lagrangian
\begin{align}
\label{eq:qlag}
\frac{\mathcal{L}}{\sqrt{g}}=&K(X)+G_4(X)R\nonumber\\&+G_{4,\,X}\left((\Box\phi)^2-\nabla_\mu\nabla_\nu\phi\nabla^\mu\nabla^\nu\phi\right),
\end{align}
where $X=-(\partial\phi)^2/2$, $K(X)=X$, and
\begin{align}
G_4(X)=\frac{\mpl^2}{2}+2c_0\frac{\phi}{\mpl}+2\frac{c_4}{\Lambda_4^6}X^2.
\end{align}
(We have chosen the notation to match that commonly used in the literature.) The free parameters $c_0$ (often called $\alpha$ or $\beta$ elsewhere in the literature) and $c_4$ parameterize the strength of the coupling to matter and the strength of the new interaction respectively. The speed of gravitational waves in this theory is given by \cite{Brax:2015dma} 
\begin{equation}
\left\vert \frac{c_T^2-c^2}{c^2}\right\vert=\left\vert\frac{4c_4x^2}{1-3c_4x^2}\right\vert<\ctlimtt
\end{equation}
imposing the LIGO/VIRGO-Fermi bound. The parameter $x=\dot{\phi}/(H\mpl)$ where $\dot{\phi}$ is the time-derivative of the scalar (using cosmic time) encodes information about the cosmology of the galileon. Indeed, one has \cite{Appleby:2011aa} 
\begin{equation}\label{eq:omega}
\Omega_\phi=c_0 x^2+\frac{x^2}{6}+\frac{15c_4x^2}{2}.
\end{equation}
where $\Omega_\phi$ is the density parameter for galileons. 
The speed of gravitational waves therefore constrains a combination of $c_4$ and the cosmology of the galileon. 

On smaller scales, reference \cite{Sakstein:2017bws} has recently obtained new bounds on the parameters $c_0$ and $c_4$ using the lack of SEP violations predicted in these theories \cite{Hui:2012qt,Hui:2012jb}. Black holes in galileon theories have no scalar hair and therefore do not couple to external fields. Non-relativistic baryons do couple to galileon fields, and therefore black holes and baryons fall at different rates in external gravitational fields, signifying a violation of the SEP. The acceleration of a satellite galaxy infalling towards the center of a cluster would have a sub-dominant galileon component not felt by its central supermassive black hole (SMBH). This would cause the SMBH to lag behind the rest of the galaxy and become offset from the center by an observable amount ($\mathcal{O}(\textrm{kpc})$) given by the distance where the missing galileon component is balanced by the restoring force from the baryons left at the center. Using the techniques of \cite{Asvathaman:2015nna} applied to the galaxy M87 (located in the Virgo cluster), \cite{Sakstein:2017bws} were able to place strong constraints on $c_0$\footnote{When applied to cosmological galileons, the effective coupling to matter is a combination of $c_0$ and the cosmological parameters \cite{Kimura:2011dc,Koyama:2013paa}, and so it is this combination that is constrained. We account for this in the present work.} and $c_4$.

\begin{figure}
\includegraphics[width=0.45\textwidth]{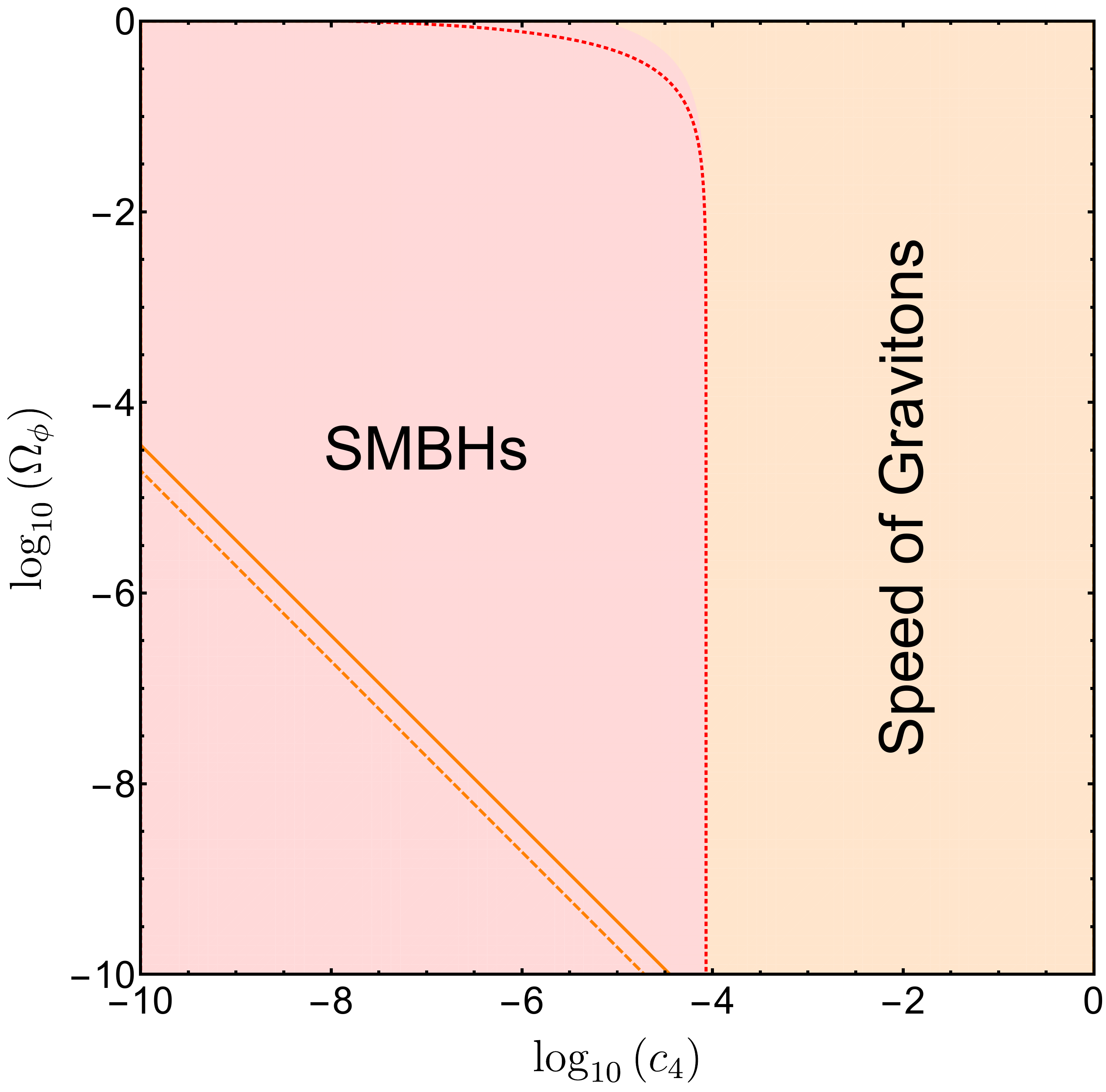}
\caption{Constraints in the $c_4$--$\Omega_\phi$ plane coming from the near equivalence of the speed of gravitons and photons (orange) and the lack of an offset supermassive black hole in M87 (red). The shaded regions correspond to $c_0=1$, and we indicate the extent of the graviton speed constraint using the solid orange line. Also shown using the dashed and {dotted} lines are the equivalent regions for $c_0=3$ from the {LIGO-Fermi and SMBH bounds respectively}.  }\label{fig:quartic}
\end{figure}

Taken together, the SMBH and LIGO/VRIGO-Fermi  constraints allow one to constrain the cosmological contribution of the quartic galileon to the Universe's energy budget using equation \eqref{eq:omega}. In Figure \ref{fig:quartic} we show the corresponding constraints in the $c_4$--$\Omega_\phi$ plane for representative values of $c_0\sim\mathcal{O}(1)$\footnote{Order-unity matter couplings are considered natural in scalar-tensor theories. One can write $c_0\phi/\mpl=\phi/M$ with $M=\mpl/c_0$ being the relevant mass-scale for the interaction. $c_0\ll1\Rightarrow M\gg\mpl$ and the theory has a trans-Planckian mass-scale whereas $c_0\gg1\Rightarrow M\ll\mpl$ so that there is a low cut-off for the effective field theory (at least n\"{a}ively, the Vainshtein mechanism alters the cut-off for the theory and the quantum properties of galileons are still uncertain \cite{Kaloper:2014vqa,Keltner:2015xda,deRham:2017avq,deRham:2017imi,Millington:2017sea}). }. The LIGO/VIRGO-Fermi bounds constrain large values of $c_4$, which correspond to large differences in the speed of photons and gravitons as well as strong screening; whereas SMBH constrains small $c_4$, where galaxy clusters are less screened and the speed of photons and gravitons are similar. The speed of gravitons therefore rules out larger values of $\Omega_\phi$ at large $c_4$ while SMBH constraints rule out lower values at low $c_4$. The two constraints are therefore complementary, and the combination rules out all galileon cosmologies except those where $\Omega_\phi\ll1$. In particular, the combination of SMBH and LIGO/VIRGO-Fermi constraints rules out regions where the galileon is more important than radiation in the late-time Universe. Clearly, the galileon can have little to nothing to say about dark energy. One could potentially avoid these harsh restrictions by adding other terms such as a cubic galileon (which itself is heavily constrained by a prediction of a too large integrated Sachs-Wolfe effect \cite{Barreira:2012kk,Renk:2017rzu}) or a quintic galileon (which typically destabilizes Vainshtein screening \cite{Kimura:2011dc,Koyama:2013paa}). We will not do so here because, ultimately, one is simply adding more parameters to the theory, in which case there are bound to be tunings that can circumvent constraints.

An alternative to the covariant quartic galileon is the beyond Horndeski quartic galileon\footnote{With Lagrangian $\mathcal{L}/\sqrt{-g}=\mpl^2R/2+X+2c_4X^2/\Lambda_4^6[(\Box\phi)^2-\nabla_\mu\nabla_\nu\phi\nabla^\mu\nabla^\nu\phi]$.}. This model gives an identical cosmology to the model studied above and has an identical expression for $c_T^2$ \cite{Kase:2014yya}. Therefore, this model is also tightly constrained. Going beyond this, a large portion of Horndeski and BH models are now excluded as dark energy candidates, as are several more complicated theories such as vector-tensor and degenerate higher-order scalar-tensor theories \cite{Baker:2017hug,Ezquiaga:2017ekz,Creminelli:2017sry}. We emphasize that models such as Einstein-dilaton-Gauss-Bonnet that do not lead an accelerating universe, of interest for example for deviations detectable via black hole tests \cite{Kanti:1995vq}, are still allowed.

Finally, we consider one additional quantity that can be bounded by the LIGO/VIRGO-Fermi  observation: the disformal coupling to photons. Disformal couplings refer to derivative couplings of the scalar to a matter species $i$ via the metric
\begin{equation}
\tilde{g}_{\mu\nu}^{(i)}=g_{\mu\nu}+\frac{\partial_\mu\phi\partial_\nu\phi}{\mathcal{M}_i^4}.
\end{equation}
If $\mathcal{M}_i=\mathcal{M}$ then the field couples to all species universally and there is no violation of the equivalence principle. But there is no \textit{a priori} reason for one to expect this to be the case and, in particular, reference \cite{vandeBruck:2016cnh} have investigated the effects of having different disformal couplings to photons and baryons. In the simplest case where there is no disformal coupling to matter\footnote{More complicated theories where the cosmic acceleration arises from disformal couplings \cite{Berezhiani:2016dne} are also constrained as photons travel slower than gravitons in these theories.}, the speed of photons is given by
\begin{equation}
\frac{c_\gamma^2}{c^2}=1-\frac{\dot{\phi}^2}{\mathcal{M}_\gamma^4},
\end{equation}
and so the LIGO/VIRGO-Fermi bound implies that ${\dot{\phi}^2}/{\mathcal{M}_\gamma^4}\lsim\ctlimtt$. Cosmologically-relevant scalars typically have $\dot{\phi}\sim H_0\mpl$ \cite{Copeland:2006wr}, which implies that $\mathcal{M}_\gamma\gsim 10$ MeV. This is stronger than the constraint inferred from the absence of any vacuum \v{C}erenkov radiation observed at LEP \cite{Hohensee:2009zk} and is comparable with the bound from constraints on energy-loss by the Primakov process in the Sun. It is superseded by the same constraints coming from horizontal branch stars, which give $M_\gamma\gsim100$ MeV \cite{Brax:2014vva}, although our bound is free from the degeneracies of stellar physics (such as metallicity). At this level, the disformal coupling to photons can have no significant effect on the cosmic evolution of the scalar \cite{Sakstein:2015jca}.

To summarize, in this letter we have highlighted three important consequences of the observation of gravitational waves and an optical counterpart from the binary neutron star merger GW170817 \cite{TheLIGOScientific:2017qsa,Monitor:2017mdv,GBM:2017lvd} for cosmological scalar-tensor theories. The close arrival time (less than a minute) constrains the speed of gravitons and photons to differ by at most one part in $\ctlim$. For beyond Horndeski theories, a very general framework for constructing pathology-free dark energy models, one of the five functions that describes the cosmology of these theories ($\alpha_T=c_T^2/c^2-1$) is now known to be negligible. Furthermore, this implies that two of the other functions, $\alpha_B$ and $\alpha_H$, can be constrained using astrophysical tests. We have presented these constraints here for the first time. Second, combining the LIGO/VIRGO-Fermi bound with separate bounds coming from the lack of any strong equivalence principle violations by the central supermassive black hole in M87, we have shown that the covariant quartic galileon, a common paradigm for modified gravity as dark energy, must be cosmologically irrelevant. Finally, we have constrained the disformal coupling to photons, and shown that this can play no significant role in the cosmological evolution of scalar fields.  

{\bf Acknowledgements:} We are grateful for discussions with Tessa Baker, Jeremy Heyl, Justin Khoury, Kazuya Koyama, Mark Trodden, and Filippo Vernizzi. BJ is supported in part by the US Department of Energy grant DE-SC0007901. JS is supported by funds provided to the Center for Particle Cosmology by the University of Pennsylvania.

\bibliography{ref}

\begin{thebibliography}{82}%
\makeatletter
\providecommand \@ifxundefined [1]{%
 \@ifx{#1\undefined}
}%
\providecommand \@ifnum [1]{%
 \ifnum #1\expandafter \@firstoftwo
 \else \expandafter \@secondoftwo
 \fi
}%
\providecommand \@ifx [1]{%
 \ifx #1\expandafter \@firstoftwo
 \else \expandafter \@secondoftwo
 \fi
}%
\providecommand \natexlab [1]{#1}%
\providecommand \enquote  [1]{``#1''}%
\providecommand \bibnamefont  [1]{#1}%
\providecommand \bibfnamefont [1]{#1}%
\providecommand \citenamefont [1]{#1}%
\providecommand \href@noop [0]{\@secondoftwo}%
\providecommand \href [0]{\begingroup \@sanitize@url \@href}%
\providecommand \@href[1]{\@@startlink{#1}\@@href}%
\providecommand \@@href[1]{\endgroup#1\@@endlink}%
\providecommand \@sanitize@url [0]{\catcode `\\12\catcode `\$12\catcode
  `\&12\catcode `\#12\catcode `\^12\catcode `\_12\catcode `\%12\relax}%
\providecommand \@@startlink[1]{}%
\providecommand \@@endlink[0]{}%
\providecommand \url  [0]{\begingroup\@sanitize@url \@url }%
\providecommand \@url [1]{\endgroup\@href {#1}{\urlprefix }}%
\providecommand \urlprefix  [0]{URL }%
\providecommand \Eprint [0]{\href }%
\providecommand \doibase [0]{http://dx.doi.org/}%
\providecommand \selectlanguage [0]{\@gobble}%
\providecommand \bibinfo  [0]{\@secondoftwo}%
\providecommand \bibfield  [0]{\@secondoftwo}%
\providecommand \translation [1]{[#1]}%
\providecommand \BibitemOpen [0]{}%
\providecommand \bibitemStop [0]{}%
\providecommand \bibitemNoStop [0]{.\EOS\space}%
\providecommand \EOS [0]{\spacefactor3000\relax}%
\providecommand \BibitemShut  [1]{\csname bibitem#1\endcsname}%
\let\auto@bib@innerbib\@empty
\bibitem [{\citenamefont {Clifton}\ \emph {et~al.}(2012)\citenamefont
  {Clifton}, \citenamefont {Ferreira}, \citenamefont {Padilla},\ and\
  \citenamefont {Skordis}}]{Clifton:2011jh}%
  \BibitemOpen
  \bibfield  {author} {\bibinfo {author} {\bibfnamefont {T.}~\bibnamefont
  {Clifton}}, \bibinfo {author} {\bibfnamefont {P.~G.}\ \bibnamefont
  {Ferreira}}, \bibinfo {author} {\bibfnamefont {A.}~\bibnamefont {Padilla}}, \
  and\ \bibinfo {author} {\bibfnamefont {C.}~\bibnamefont {Skordis}},\ }\href
  {\doibase 10.1016/j.physrep.2012.01.001} {\bibfield  {journal} {\bibinfo
  {journal} {Phys. Rept.}\ }\textbf {\bibinfo {volume} {513}},\ \bibinfo
  {pages} {1} (\bibinfo {year} {2012})},\ \Eprint
  {http://arxiv.org/abs/1106.2476} {arXiv:1106.2476 [astro-ph.CO]} \BibitemShut
  {NoStop}%
\bibitem [{\citenamefont {Joyce}\ \emph {et~al.}(2015)\citenamefont {Joyce},
  \citenamefont {Jain}, \citenamefont {Khoury},\ and\ \citenamefont
  {Trodden}}]{Joyce:2014kja}%
  \BibitemOpen
  \bibfield  {author} {\bibinfo {author} {\bibfnamefont {A.}~\bibnamefont
  {Joyce}}, \bibinfo {author} {\bibfnamefont {B.}~\bibnamefont {Jain}},
  \bibinfo {author} {\bibfnamefont {J.}~\bibnamefont {Khoury}}, \ and\ \bibinfo
  {author} {\bibfnamefont {M.}~\bibnamefont {Trodden}},\ }\href {\doibase
  10.1016/j.physrep.2014.12.002} {\bibfield  {journal} {\bibinfo  {journal}
  {Phys. Rept.}\ }\textbf {\bibinfo {volume} {568}},\ \bibinfo {pages} {1}
  (\bibinfo {year} {2015})},\ \Eprint {http://arxiv.org/abs/1407.0059}
  {arXiv:1407.0059 [astro-ph.CO]} \BibitemShut {NoStop}%
\bibitem [{\citenamefont {Lombriser}(2014)}]{Lombriser:2014dua}%
  \BibitemOpen
  \bibfield  {author} {\bibinfo {author} {\bibfnamefont {L.}~\bibnamefont
  {Lombriser}},\ }\href {\doibase 10.1002/andp.201400058} {\bibfield  {journal}
  {\bibinfo  {journal} {Annalen Phys.}\ }\textbf {\bibinfo {volume} {526}},\
  \bibinfo {pages} {259} (\bibinfo {year} {2014})},\ \Eprint
  {http://arxiv.org/abs/1403.4268} {arXiv:1403.4268 [astro-ph.CO]} \BibitemShut
  {NoStop}%
\bibitem [{\citenamefont {Bull}\ \emph {et~al.}(2016)\citenamefont {Bull} \emph
  {et~al.}}]{Bull:2015stt}%
  \BibitemOpen
  \bibfield  {author} {\bibinfo {author} {\bibfnamefont {P.}~\bibnamefont
  {Bull}} \emph {et~al.},\ }\href {\doibase 10.1016/j.dark.2016.02.001}
  {\bibfield  {journal} {\bibinfo  {journal} {Phys. Dark Univ.}\ }\textbf
  {\bibinfo {volume} {12}},\ \bibinfo {pages} {56} (\bibinfo {year} {2016})},\
  \Eprint {http://arxiv.org/abs/1512.05356} {arXiv:1512.05356 [astro-ph.CO]}
  \BibitemShut {NoStop}%
\bibitem [{\citenamefont {Koyama}(2016)}]{Koyama:2015vza}%
  \BibitemOpen
  \bibfield  {author} {\bibinfo {author} {\bibfnamefont {K.}~\bibnamefont
  {Koyama}},\ }\href {\doibase 10.1088/0034-4885/79/4/046902} {\bibfield
  {journal} {\bibinfo  {journal} {Rept. Prog. Phys.}\ }\textbf {\bibinfo
  {volume} {79}},\ \bibinfo {pages} {046902} (\bibinfo {year} {2016})},\
  \Eprint {http://arxiv.org/abs/1504.04623} {arXiv:1504.04623 [astro-ph.CO]}
  \BibitemShut {NoStop}%
\bibitem [{\citenamefont {de~Rham}\ \emph
  {et~al.}(2017{\natexlab{a}})\citenamefont {de~Rham}, \citenamefont {Deskins},
  \citenamefont {Tolley},\ and\ \citenamefont {Zhou}}]{deRham:2016nuf}%
  \BibitemOpen
  \bibfield  {author} {\bibinfo {author} {\bibfnamefont {C.}~\bibnamefont
  {de~Rham}}, \bibinfo {author} {\bibfnamefont {J.~T.}\ \bibnamefont
  {Deskins}}, \bibinfo {author} {\bibfnamefont {A.~J.}\ \bibnamefont {Tolley}},
  \ and\ \bibinfo {author} {\bibfnamefont {S.-Y.}\ \bibnamefont {Zhou}},\
  }\href {\doibase 10.1103/RevModPhys.89.025004} {\bibfield  {journal}
  {\bibinfo  {journal} {Rev. Mod. Phys.}\ }\textbf {\bibinfo {volume} {89}},\
  \bibinfo {pages} {025004} (\bibinfo {year} {2017}{\natexlab{a}})},\ \Eprint
  {http://arxiv.org/abs/1606.08462} {arXiv:1606.08462 [astro-ph.CO]}
  \BibitemShut {NoStop}%
\bibitem [{\citenamefont {Burrage}\ and\ \citenamefont
  {Sakstein}(2016)}]{Burrage:2016bwy}%
  \BibitemOpen
  \bibfield  {author} {\bibinfo {author} {\bibfnamefont {C.}~\bibnamefont
  {Burrage}}\ and\ \bibinfo {author} {\bibfnamefont {J.}~\bibnamefont
  {Sakstein}},\ }\href {\doibase 10.1088/1475-7516/2016/11/045} {\bibfield
  {journal} {\bibinfo  {journal} {JCAP}\ }\textbf {\bibinfo {volume} {1611}},\
  \bibinfo {pages} {045} (\bibinfo {year} {2016})},\ \Eprint
  {http://arxiv.org/abs/1609.01192} {arXiv:1609.01192 [astro-ph.CO]}
  \BibitemShut {NoStop}%
\bibitem [{\citenamefont {Burrage}\ and\ \citenamefont
  {Sakstein}(2017)}]{Burrage:2017qrf}%
  \BibitemOpen
  \bibfield  {author} {\bibinfo {author} {\bibfnamefont {C.}~\bibnamefont
  {Burrage}}\ and\ \bibinfo {author} {\bibfnamefont {J.}~\bibnamefont
  {Sakstein}},\ }\href@noop {} {\  (\bibinfo {year} {2017})},\ \Eprint
  {http://arxiv.org/abs/1709.09071} {arXiv:1709.09071 [astro-ph.CO]}
  \BibitemShut {NoStop}%
\bibitem [{\citenamefont {Sakstein}(2017)}]{Sakstein:2017pqi}%
  \BibitemOpen
  \bibfield  {author} {\bibinfo {author} {\bibfnamefont {J.}~\bibnamefont
  {Sakstein}},\ }\href@noop {} {\  (\bibinfo {year} {2017})},\ \Eprint
  {http://arxiv.org/abs/1710.03156} {arXiv:1710.03156 [astro-ph.CO]}
  \BibitemShut {NoStop}%
\bibitem [{\citenamefont {Gleyzes}\ \emph
  {et~al.}(2015{\natexlab{a}})\citenamefont {Gleyzes}, \citenamefont
  {Langlois}, \citenamefont {Piazza},\ and\ \citenamefont
  {Vernizzi}}]{Gleyzes:2014dya}%
  \BibitemOpen
  \bibfield  {author} {\bibinfo {author} {\bibfnamefont {J.}~\bibnamefont
  {Gleyzes}}, \bibinfo {author} {\bibfnamefont {D.}~\bibnamefont {Langlois}},
  \bibinfo {author} {\bibfnamefont {F.}~\bibnamefont {Piazza}}, \ and\ \bibinfo
  {author} {\bibfnamefont {F.}~\bibnamefont {Vernizzi}},\ }\href {\doibase
  10.1103/PhysRevLett.114.211101} {\bibfield  {journal} {\bibinfo  {journal}
  {Phys. Rev. Lett.}\ }\textbf {\bibinfo {volume} {114}},\ \bibinfo {pages}
  {211101} (\bibinfo {year} {2015}{\natexlab{a}})},\ \Eprint
  {http://arxiv.org/abs/1404.6495} {arXiv:1404.6495 [hep-th]} \BibitemShut
  {NoStop}%
\bibitem [{\citenamefont {Gleyzes}\ \emph
  {et~al.}(2015{\natexlab{b}})\citenamefont {Gleyzes}, \citenamefont
  {Langlois}, \citenamefont {Piazza},\ and\ \citenamefont
  {Vernizzi}}]{Gleyzes:2014qga}%
  \BibitemOpen
  \bibfield  {author} {\bibinfo {author} {\bibfnamefont {J.}~\bibnamefont
  {Gleyzes}}, \bibinfo {author} {\bibfnamefont {D.}~\bibnamefont {Langlois}},
  \bibinfo {author} {\bibfnamefont {F.}~\bibnamefont {Piazza}}, \ and\ \bibinfo
  {author} {\bibfnamefont {F.}~\bibnamefont {Vernizzi}},\ }\href {\doibase
  10.1088/1475-7516/2015/02/018} {\bibfield  {journal} {\bibinfo  {journal}
  {JCAP}\ }\textbf {\bibinfo {volume} {1502}},\ \bibinfo {pages} {018}
  (\bibinfo {year} {2015}{\natexlab{b}})},\ \Eprint
  {http://arxiv.org/abs/1408.1952} {arXiv:1408.1952 [astro-ph.CO]} \BibitemShut
  {NoStop}%
\bibitem [{\citenamefont {Khoury}\ and\ \citenamefont
  {Weltman}(2004)}]{Khoury:2003rn}%
  \BibitemOpen
  \bibfield  {author} {\bibinfo {author} {\bibfnamefont {J.}~\bibnamefont
  {Khoury}}\ and\ \bibinfo {author} {\bibfnamefont {A.}~\bibnamefont
  {Weltman}},\ }\href {\doibase 10.1103/PhysRevD.69.044026} {\bibfield
  {journal} {\bibinfo  {journal} {Phys. Rev.}\ }\textbf {\bibinfo {volume}
  {D69}},\ \bibinfo {pages} {044026} (\bibinfo {year} {2004})},\ \Eprint
  {http://arxiv.org/abs/astro-ph/0309411} {arXiv:astro-ph/0309411 [astro-ph]}
  \BibitemShut {NoStop}%
\bibitem [{\citenamefont {Nicolis}\ \emph {et~al.}(2009)\citenamefont
  {Nicolis}, \citenamefont {Rattazzi},\ and\ \citenamefont
  {Trincherini}}]{Nicolis:2008in}%
  \BibitemOpen
  \bibfield  {author} {\bibinfo {author} {\bibfnamefont {A.}~\bibnamefont
  {Nicolis}}, \bibinfo {author} {\bibfnamefont {R.}~\bibnamefont {Rattazzi}}, \
  and\ \bibinfo {author} {\bibfnamefont {E.}~\bibnamefont {Trincherini}},\
  }\href {\doibase 10.1103/PhysRevD.79.064036} {\bibfield  {journal} {\bibinfo
  {journal} {Phys. Rev.}\ }\textbf {\bibinfo {volume} {D79}},\ \bibinfo {pages}
  {064036} (\bibinfo {year} {2009})},\ \Eprint {http://arxiv.org/abs/0811.2197}
  {arXiv:0811.2197 [hep-th]} \BibitemShut {NoStop}%
\bibitem [{\citenamefont {Alonso}\ \emph {et~al.}(2017)\citenamefont {Alonso},
  \citenamefont {Bellini}, \citenamefont {Ferreira},\ and\ \citenamefont
  {Zumalacárregui}}]{Alonso:2016suf}%
  \BibitemOpen
  \bibfield  {author} {\bibinfo {author} {\bibfnamefont {D.}~\bibnamefont
  {Alonso}}, \bibinfo {author} {\bibfnamefont {E.}~\bibnamefont {Bellini}},
  \bibinfo {author} {\bibfnamefont {P.~G.}\ \bibnamefont {Ferreira}}, \ and\
  \bibinfo {author} {\bibfnamefont {M.}~\bibnamefont {Zumalacárregui}},\ }\href
  {\doibase 10.1103/PhysRevD.95.063502} {\bibfield  {journal} {\bibinfo
  {journal} {Phys. Rev.}\ }\textbf {\bibinfo {volume} {D95}},\ \bibinfo {pages}
  {063502} (\bibinfo {year} {2017})},\ \Eprint
  {http://arxiv.org/abs/1610.09290} {arXiv:1610.09290 [astro-ph.CO]}
  \BibitemShut {NoStop}%
\bibitem [{\citenamefont {De~Felice}\ and\ \citenamefont
  {Tsujikawa}(2012)}]{DeFelice:2011bh}%
  \BibitemOpen
  \bibfield  {author} {\bibinfo {author} {\bibfnamefont {A.}~\bibnamefont
  {De~Felice}}\ and\ \bibinfo {author} {\bibfnamefont {S.}~\bibnamefont
  {Tsujikawa}},\ }\href {\doibase 10.1088/1475-7516/2012/02/007} {\bibfield
  {journal} {\bibinfo  {journal} {JCAP}\ }\textbf {\bibinfo {volume} {1202}},\
  \bibinfo {pages} {007} (\bibinfo {year} {2012})},\ \Eprint
  {http://arxiv.org/abs/1110.3878} {arXiv:1110.3878 [gr-qc]} \BibitemShut
  {NoStop}%
\bibitem [{\citenamefont {Bellini}\ and\ \citenamefont
  {Sawicki}(2014)}]{Bellini:2014fua}%
  \BibitemOpen
  \bibfield  {author} {\bibinfo {author} {\bibfnamefont {E.}~\bibnamefont
  {Bellini}}\ and\ \bibinfo {author} {\bibfnamefont {I.}~\bibnamefont
  {Sawicki}},\ }\href {\doibase 10.1088/1475-7516/2014/07/050} {\bibfield
  {journal} {\bibinfo  {journal} {JCAP}\ }\textbf {\bibinfo {volume} {1407}},\
  \bibinfo {pages} {050} (\bibinfo {year} {2014})},\ \Eprint
  {http://arxiv.org/abs/1404.3713} {arXiv:1404.3713 [astro-ph.CO]} \BibitemShut
  {NoStop}%
\bibitem [{\citenamefont {Lombriser}\ and\ \citenamefont
  {Taylor}(2016)}]{Lombriser:2015sxa}%
  \BibitemOpen
  \bibfield  {author} {\bibinfo {author} {\bibfnamefont {L.}~\bibnamefont
  {Lombriser}}\ and\ \bibinfo {author} {\bibfnamefont {A.}~\bibnamefont
  {Taylor}},\ }\href {\doibase 10.1088/1475-7516/2016/03/031} {\bibfield
  {journal} {\bibinfo  {journal} {JCAP}\ }\textbf {\bibinfo {volume} {1603}},\
  \bibinfo {pages} {031} (\bibinfo {year} {2016})},\ \Eprint
  {http://arxiv.org/abs/1509.08458} {arXiv:1509.08458 [astro-ph.CO]}
  \BibitemShut {NoStop}%
\bibitem [{\citenamefont {Bettoni}\ \emph {et~al.}(2017)\citenamefont
  {Bettoni}, \citenamefont {Ezquiaga}, \citenamefont {Hinterbichler},\ and\
  \citenamefont {Zumalacárregui}}]{Bettoni:2016mij}%
  \BibitemOpen
  \bibfield  {author} {\bibinfo {author} {\bibfnamefont {D.}~\bibnamefont
  {Bettoni}}, \bibinfo {author} {\bibfnamefont {J.~M.}\ \bibnamefont
  {Ezquiaga}}, \bibinfo {author} {\bibfnamefont {K.}~\bibnamefont
  {Hinterbichler}}, \ and\ \bibinfo {author} {\bibfnamefont {M.}~\bibnamefont
  {Zumalacárregui}},\ }\href {\doibase 10.1103/PhysRevD.95.084029} {\bibfield
  {journal} {\bibinfo  {journal} {Phys. Rev.}\ }\textbf {\bibinfo {volume}
  {D95}},\ \bibinfo {pages} {084029} (\bibinfo {year} {2017})},\ \Eprint
  {http://arxiv.org/abs/1608.01982} {arXiv:1608.01982 [gr-qc]} \BibitemShut
  {NoStop}%
\bibitem [{\citenamefont {Abbott}\ \emph
  {et~al.}(2017{\natexlab{a}})\citenamefont {Abbott} \emph
  {et~al.}}]{TheLIGOScientific:2017qsa}%
  \BibitemOpen
  \bibfield  {author} {\bibinfo {author} {\bibfnamefont {B.~P.}\ \bibnamefont
  {Abbott}} \emph {et~al.} (\bibinfo {collaboration} {Virgo, LIGO
  Scientific}),\ }\href {\doibase 10.1103/PhysRevLett.119.161101} {\bibfield
  {journal} {\bibinfo  {journal} {Phys. Rev. Lett.}\ }\textbf {\bibinfo
  {volume} {119}},\ \bibinfo {pages} {161101} (\bibinfo {year}
  {2017}{\natexlab{a}})},\ \Eprint {http://arxiv.org/abs/1710.05832}
  {arXiv:1710.05832 [gr-qc]} \BibitemShut {NoStop}%
\bibitem [{\citenamefont {Abbott}\ \emph
  {et~al.}(2017{\natexlab{b}})\citenamefont {Abbott} \emph
  {et~al.}}]{Monitor:2017mdv}%
  \BibitemOpen
  \bibfield  {author} {\bibinfo {author} {\bibfnamefont {B.~P.}\ \bibnamefont
  {Abbott}} \emph {et~al.} (\bibinfo {collaboration} {Virgo, Fermi-GBM,
  INTEGRAL, LIGO Scientific}),\ }\href {\doibase 10.3847/2041-8213/aa920c}
  {\bibfield  {journal} {\bibinfo  {journal} {Astrophys. J.}\ }\textbf
  {\bibinfo {volume} {848}},\ \bibinfo {pages} {L13} (\bibinfo {year}
  {2017}{\natexlab{b}})},\ \Eprint {http://arxiv.org/abs/1710.05834}
  {arXiv:1710.05834 [astro-ph.HE]} \BibitemShut {NoStop}%
\bibitem [{\citenamefont {GBM}\ \emph {et~al.}(2017)\citenamefont {GBM} \emph
  {et~al.}}]{GBM:2017lvd}%
  \BibitemOpen
  \bibfield  {author} {\bibinfo {author} {\bibfnamefont {F.}~\bibnamefont
  {GBM}} \emph {et~al.} (\bibinfo {collaboration} {Transient Robotic
  Observatory of the South, The 1M2H Team, Virgo, Euro VLBI Team, The VINROUGE,
  IceCube, GRAvitational Wave Inaf TeAm, CAASTRO s, Pi of the Sky, MASTER, The
  Swift, The CALET, AstroSat Cadmium Zinc Telluride Imager Team, The BOOTES,
  The Fermi Large Area Telescope, IKI-GW Follow-up, The DLT40, HAWC, AGILE
  Team, University The Chandra Team at McGill, ALMA, NuSTAR s, LIGO Scientific,
  the DES, LOFAR, IPN, The Insight-Hxmt, The MAXI Team, ANTARES, KU, The Dark
  Energy Camera GW-EM, The Pierre Auger, H. E. S. S.}),\ }\href {\doibase
  10.3847/2041-8213/aa91c9} {\bibfield  {journal} {\bibinfo  {journal}
  {Astrophys. J.}\ }\textbf {\bibinfo {volume} {848}},\ \bibinfo {pages} {L12}
  (\bibinfo {year} {2017})},\ \Eprint {http://arxiv.org/abs/1710.05833}
  {arXiv:1710.05833 [astro-ph.HE]} \BibitemShut {NoStop}%
\bibitem [{\citenamefont {Moore}\ and\ \citenamefont
  {Nelson}(2001)}]{Moore:2001bv}%
  \BibitemOpen
  \bibfield  {author} {\bibinfo {author} {\bibfnamefont {G.~D.}\ \bibnamefont
  {Moore}}\ and\ \bibinfo {author} {\bibfnamefont {A.~E.}\ \bibnamefont
  {Nelson}},\ }\href {\doibase 10.1088/1126-6708/2001/09/023} {\bibfield
  {journal} {\bibinfo  {journal} {JHEP}\ }\textbf {\bibinfo {volume} {09}},\
  \bibinfo {pages} {023} (\bibinfo {year} {2001})},\ \Eprint
  {http://arxiv.org/abs/hep-ph/0106220} {arXiv:hep-ph/0106220 [hep-ph]}
  \BibitemShut {NoStop}%
\bibitem [{\citenamefont {Hohensee}\ \emph {et~al.}(2009)\citenamefont
  {Hohensee}, \citenamefont {Lehnert}, \citenamefont {Phillips},\ and\
  \citenamefont {Walsworth}}]{Hohensee:2009zk}%
  \BibitemOpen
  \bibfield  {author} {\bibinfo {author} {\bibfnamefont {M.~A.}\ \bibnamefont
  {Hohensee}}, \bibinfo {author} {\bibfnamefont {R.}~\bibnamefont {Lehnert}},
  \bibinfo {author} {\bibfnamefont {D.~F.}\ \bibnamefont {Phillips}}, \ and\
  \bibinfo {author} {\bibfnamefont {R.~L.}\ \bibnamefont {Walsworth}},\ }\href
  {\doibase 10.1103/PhysRevLett.102.170402} {\bibfield  {journal} {\bibinfo
  {journal} {Phys. Rev. Lett.}\ }\textbf {\bibinfo {volume} {102}},\ \bibinfo
  {pages} {170402} (\bibinfo {year} {2009})},\ \Eprint
  {http://arxiv.org/abs/0904.2031} {arXiv:0904.2031 [hep-ph]} \BibitemShut
  {NoStop}%
\bibitem [{\citenamefont {Gleyzes}\ \emph
  {et~al.}(2015{\natexlab{c}})\citenamefont {Gleyzes}, \citenamefont
  {Langlois},\ and\ \citenamefont {Vernizzi}}]{Gleyzes:2014rba}%
  \BibitemOpen
  \bibfield  {author} {\bibinfo {author} {\bibfnamefont {J.}~\bibnamefont
  {Gleyzes}}, \bibinfo {author} {\bibfnamefont {D.}~\bibnamefont {Langlois}}, \
  and\ \bibinfo {author} {\bibfnamefont {F.}~\bibnamefont {Vernizzi}},\ }\href
  {\doibase 10.1142/S021827181443010X} {\bibfield  {journal} {\bibinfo
  {journal} {Int. J. Mod. Phys.}\ }\textbf {\bibinfo {volume} {D23}},\ \bibinfo
  {pages} {1443010} (\bibinfo {year} {2015}{\natexlab{c}})},\ \Eprint
  {http://arxiv.org/abs/1411.3712} {arXiv:1411.3712 [hep-th]} \BibitemShut
  {NoStop}%
\bibitem [{\citenamefont {Creminelli}\ \emph {et~al.}(2009)\citenamefont
  {Creminelli}, \citenamefont {D'Amico}, \citenamefont {Norena},\ and\
  \citenamefont {Vernizzi}}]{Creminelli:2008wc}%
  \BibitemOpen
  \bibfield  {author} {\bibinfo {author} {\bibfnamefont {P.}~\bibnamefont
  {Creminelli}}, \bibinfo {author} {\bibfnamefont {G.}~\bibnamefont {D'Amico}},
  \bibinfo {author} {\bibfnamefont {J.}~\bibnamefont {Norena}}, \ and\ \bibinfo
  {author} {\bibfnamefont {F.}~\bibnamefont {Vernizzi}},\ }\href {\doibase
  10.1088/1475-7516/2009/02/018} {\bibfield  {journal} {\bibinfo  {journal}
  {JCAP}\ }\textbf {\bibinfo {volume} {0902}},\ \bibinfo {pages} {018}
  (\bibinfo {year} {2009})},\ \Eprint {http://arxiv.org/abs/0811.0827}
  {arXiv:0811.0827 [astro-ph]} \BibitemShut {NoStop}%
\bibitem [{\citenamefont {Gubitosi}\ \emph {et~al.}(2013)\citenamefont
  {Gubitosi}, \citenamefont {Piazza},\ and\ \citenamefont
  {Vernizzi}}]{Gubitosi:2012hu}%
  \BibitemOpen
  \bibfield  {author} {\bibinfo {author} {\bibfnamefont {G.}~\bibnamefont
  {Gubitosi}}, \bibinfo {author} {\bibfnamefont {F.}~\bibnamefont {Piazza}}, \
  and\ \bibinfo {author} {\bibfnamefont {F.}~\bibnamefont {Vernizzi}},\ }\href
  {\doibase 10.1088/1475-7516/2013/02/032} {\bibfield  {journal} {\bibinfo
  {journal} {JCAP}\ }\textbf {\bibinfo {volume} {1302}},\ \bibinfo {pages}
  {032} (\bibinfo {year} {2013})},\ \bibinfo {note} {[JCAP1302,032(2013)]},\
  \Eprint {http://arxiv.org/abs/1210.0201} {arXiv:1210.0201 [hep-th]}
  \BibitemShut {NoStop}%
\bibitem [{\citenamefont {Bloomfield}\ \emph {et~al.}(2013)\citenamefont
  {Bloomfield}, \citenamefont {Flanagan}, \citenamefont {Park},\ and\
  \citenamefont {Watson}}]{Bloomfield:2012ff}%
  \BibitemOpen
  \bibfield  {author} {\bibinfo {author} {\bibfnamefont {J.~K.}\ \bibnamefont
  {Bloomfield}}, \bibinfo {author} {\bibfnamefont {É.~É.}\ \bibnamefont
  {Flanagan}}, \bibinfo {author} {\bibfnamefont {M.}~\bibnamefont {Park}}, \
  and\ \bibinfo {author} {\bibfnamefont {S.}~\bibnamefont {Watson}},\ }\href
  {\doibase 10.1088/1475-7516/2013/08/010} {\bibfield  {journal} {\bibinfo
  {journal} {JCAP}\ }\textbf {\bibinfo {volume} {1308}},\ \bibinfo {pages}
  {010} (\bibinfo {year} {2013})},\ \Eprint {http://arxiv.org/abs/1211.7054}
  {arXiv:1211.7054 [astro-ph.CO]} \BibitemShut {NoStop}%
\bibitem [{\citenamefont {Bekenstein}(1993)}]{Bekenstein:1992pj}%
  \BibitemOpen
  \bibfield  {author} {\bibinfo {author} {\bibfnamefont {J.~D.}\ \bibnamefont
  {Bekenstein}},\ }\href {\doibase 10.1103/PhysRevD.48.3641} {\bibfield
  {journal} {\bibinfo  {journal} {Phys. Rev.}\ }\textbf {\bibinfo {volume}
  {D48}},\ \bibinfo {pages} {3641} (\bibinfo {year} {1993})},\ \Eprint
  {http://arxiv.org/abs/gr-qc/9211017} {arXiv:gr-qc/9211017 [gr-qc]}
  \BibitemShut {NoStop}%
\bibitem [{\citenamefont {Sakstein}(2014)}]{Sakstein:2014isa}%
  \BibitemOpen
  \bibfield  {author} {\bibinfo {author} {\bibfnamefont {J.}~\bibnamefont
  {Sakstein}},\ }\href {\doibase 10.1088/1475-7516/2014/12/012} {\bibfield
  {journal} {\bibinfo  {journal} {JCAP}\ }\textbf {\bibinfo {volume} {1412}},\
  \bibinfo {pages} {012} (\bibinfo {year} {2014})},\ \Eprint
  {http://arxiv.org/abs/1409.1734} {arXiv:1409.1734 [astro-ph.CO]} \BibitemShut
  {NoStop}%
\bibitem [{\citenamefont {Sakstein}(2015{\natexlab{a}})}]{Sakstein:2014aca}%
  \BibitemOpen
  \bibfield  {author} {\bibinfo {author} {\bibfnamefont {J.}~\bibnamefont
  {Sakstein}},\ }\href {\doibase 10.1103/PhysRevD.91.024036} {\bibfield
  {journal} {\bibinfo  {journal} {Phys. Rev.}\ }\textbf {\bibinfo {volume}
  {D91}},\ \bibinfo {pages} {024036} (\bibinfo {year} {2015}{\natexlab{a}})},\
  \Eprint {http://arxiv.org/abs/1409.7296} {arXiv:1409.7296 [astro-ph.CO]}
  \BibitemShut {NoStop}%
\bibitem [{\citenamefont {Ip}\ \emph {et~al.}(2015)\citenamefont {Ip},
  \citenamefont {Sakstein},\ and\ \citenamefont {Schmidt}}]{Ip:2015qsa}%
  \BibitemOpen
  \bibfield  {author} {\bibinfo {author} {\bibfnamefont {H.~Y.}\ \bibnamefont
  {Ip}}, \bibinfo {author} {\bibfnamefont {J.}~\bibnamefont {Sakstein}}, \ and\
  \bibinfo {author} {\bibfnamefont {F.}~\bibnamefont {Schmidt}},\ }\href
  {\doibase 10.1088/1475-7516/2015/10/051} {\bibfield  {journal} {\bibinfo
  {journal} {JCAP}\ }\textbf {\bibinfo {volume} {1510}},\ \bibinfo {pages}
  {051} (\bibinfo {year} {2015})},\ \Eprint {http://arxiv.org/abs/1507.00568}
  {arXiv:1507.00568 [gr-qc]} \BibitemShut {NoStop}%
\bibitem [{\citenamefont {Sakstein}\ and\ \citenamefont
  {Verner}(2015)}]{Sakstein:2015jca}%
  \BibitemOpen
  \bibfield  {author} {\bibinfo {author} {\bibfnamefont {J.}~\bibnamefont
  {Sakstein}}\ and\ \bibinfo {author} {\bibfnamefont {S.}~\bibnamefont
  {Verner}},\ }\href {\doibase 10.1103/PhysRevD.92.123005} {\bibfield
  {journal} {\bibinfo  {journal} {Phys. Rev.}\ }\textbf {\bibinfo {volume}
  {D92}},\ \bibinfo {pages} {123005} (\bibinfo {year} {2015})},\ \Eprint
  {http://arxiv.org/abs/1509.05679} {arXiv:1509.05679 [gr-qc]} \BibitemShut
  {NoStop}%
\bibitem [{\citenamefont {Crisostomi}\ and\ \citenamefont
  {Koyama}(2017)}]{Crisostomi:2017lbg}%
  \BibitemOpen
  \bibfield  {author} {\bibinfo {author} {\bibfnamefont {M.}~\bibnamefont
  {Crisostomi}}\ and\ \bibinfo {author} {\bibfnamefont {K.}~\bibnamefont
  {Koyama}},\ }\href@noop {} {\  (\bibinfo {year} {2017})},\ \Eprint
  {http://arxiv.org/abs/1711.06661} {arXiv:1711.06661 [astro-ph.CO]}
  \BibitemShut {NoStop}%
\bibitem [{\citenamefont {Langlois}\ \emph {et~al.}(2017)\citenamefont
  {Langlois}, \citenamefont {Saito}, \citenamefont {Yamauchi},\ and\
  \citenamefont {Noui}}]{Langlois:2017dyl}%
  \BibitemOpen
  \bibfield  {author} {\bibinfo {author} {\bibfnamefont {D.}~\bibnamefont
  {Langlois}}, \bibinfo {author} {\bibfnamefont {R.}~\bibnamefont {Saito}},
  \bibinfo {author} {\bibfnamefont {D.}~\bibnamefont {Yamauchi}}, \ and\
  \bibinfo {author} {\bibfnamefont {K.}~\bibnamefont {Noui}},\ }\href@noop {}
  {\  (\bibinfo {year} {2017})},\ \Eprint {http://arxiv.org/abs/1711.07403}
  {arXiv:1711.07403 [gr-qc]} \BibitemShut {NoStop}%
\bibitem [{\citenamefont {Dima}\ and\ \citenamefont
  {Vernizzi}(2017)}]{Dima:2017pwp}%
  \BibitemOpen
  \bibfield  {author} {\bibinfo {author} {\bibfnamefont {A.}~\bibnamefont
  {Dima}}\ and\ \bibinfo {author} {\bibfnamefont {F.}~\bibnamefont
  {Vernizzi}},\ }\href@noop {} {\  (\bibinfo {year} {2017})},\ \Eprint
  {http://arxiv.org/abs/1712.04731} {arXiv:1712.04731 [gr-qc]} \BibitemShut
  {NoStop}%
\bibitem [{\citenamefont {Beltran~Jimenez}\ \emph {et~al.}(2016)\citenamefont
  {Beltran~Jimenez}, \citenamefont {Piazza},\ and\ \citenamefont
  {Velten}}]{Jimenez:2015bwa}%
  \BibitemOpen
  \bibfield  {author} {\bibinfo {author} {\bibfnamefont {J.}~\bibnamefont
  {Beltran~Jimenez}}, \bibinfo {author} {\bibfnamefont {F.}~\bibnamefont
  {Piazza}}, \ and\ \bibinfo {author} {\bibfnamefont {H.}~\bibnamefont
  {Velten}},\ }\href {\doibase 10.1103/PhysRevLett.116.061101} {\bibfield
  {journal} {\bibinfo  {journal} {Phys. Rev. Lett.}\ }\textbf {\bibinfo
  {volume} {116}},\ \bibinfo {pages} {061101} (\bibinfo {year} {2016})},\
  \Eprint {http://arxiv.org/abs/1507.05047} {arXiv:1507.05047 [gr-qc]}
  \BibitemShut {NoStop}%
\bibitem [{\citenamefont {Kobayashi}\ \emph {et~al.}(2015)\citenamefont
  {Kobayashi}, \citenamefont {Watanabe},\ and\ \citenamefont
  {Yamauchi}}]{Kobayashi:2014ida}%
  \BibitemOpen
  \bibfield  {author} {\bibinfo {author} {\bibfnamefont {T.}~\bibnamefont
  {Kobayashi}}, \bibinfo {author} {\bibfnamefont {Y.}~\bibnamefont {Watanabe}},
  \ and\ \bibinfo {author} {\bibfnamefont {D.}~\bibnamefont {Yamauchi}},\
  }\href {\doibase 10.1103/PhysRevD.91.064013} {\bibfield  {journal} {\bibinfo
  {journal} {Phys. Rev.}\ }\textbf {\bibinfo {volume} {D91}},\ \bibinfo {pages}
  {064013} (\bibinfo {year} {2015})},\ \Eprint {http://arxiv.org/abs/1411.4130}
  {arXiv:1411.4130 [gr-qc]} \BibitemShut {NoStop}%
\bibitem [{\citenamefont {Koyama}\ and\ \citenamefont
  {Sakstein}(2015)}]{Koyama:2015oma}%
  \BibitemOpen
  \bibfield  {author} {\bibinfo {author} {\bibfnamefont {K.}~\bibnamefont
  {Koyama}}\ and\ \bibinfo {author} {\bibfnamefont {J.}~\bibnamefont
  {Sakstein}},\ }\href {\doibase 10.1103/PhysRevD.91.124066} {\bibfield
  {journal} {\bibinfo  {journal} {Phys. Rev.}\ }\textbf {\bibinfo {volume}
  {D91}},\ \bibinfo {pages} {124066} (\bibinfo {year} {2015})},\ \Eprint
  {http://arxiv.org/abs/1502.06872} {arXiv:1502.06872 [astro-ph.CO]}
  \BibitemShut {NoStop}%
\bibitem [{\citenamefont {Saito}\ \emph {et~al.}(2015)\citenamefont {Saito},
  \citenamefont {Yamauchi}, \citenamefont {Mizuno}, \citenamefont {Gleyzes},\
  and\ \citenamefont {Langlois}}]{Saito:2015fza}%
  \BibitemOpen
  \bibfield  {author} {\bibinfo {author} {\bibfnamefont {R.}~\bibnamefont
  {Saito}}, \bibinfo {author} {\bibfnamefont {D.}~\bibnamefont {Yamauchi}},
  \bibinfo {author} {\bibfnamefont {S.}~\bibnamefont {Mizuno}}, \bibinfo
  {author} {\bibfnamefont {J.}~\bibnamefont {Gleyzes}}, \ and\ \bibinfo
  {author} {\bibfnamefont {D.}~\bibnamefont {Langlois}},\ }\href {\doibase
  10.1088/1475-7516/2015/06/008} {\bibfield  {journal} {\bibinfo  {journal}
  {JCAP}\ }\textbf {\bibinfo {volume} {1506}},\ \bibinfo {pages} {008}
  (\bibinfo {year} {2015})},\ \Eprint {http://arxiv.org/abs/1503.01448}
  {arXiv:1503.01448 [gr-qc]} \BibitemShut {NoStop}%
\bibitem [{\citenamefont {Sakstein}(2015{\natexlab{b}})}]{Sakstein:2015zoa}%
  \BibitemOpen
  \bibfield  {author} {\bibinfo {author} {\bibfnamefont {J.}~\bibnamefont
  {Sakstein}},\ }\href {\doibase 10.1103/PhysRevLett.115.201101} {\bibfield
  {journal} {\bibinfo  {journal} {Phys. Rev. Lett.}\ }\textbf {\bibinfo
  {volume} {115}},\ \bibinfo {pages} {201101} (\bibinfo {year}
  {2015}{\natexlab{b}})},\ \Eprint {http://arxiv.org/abs/1510.05964}
  {arXiv:1510.05964 [astro-ph.CO]} \BibitemShut {NoStop}%
\bibitem [{\citenamefont {Sakstein}(2015{\natexlab{c}})}]{Sakstein:2015aac}%
  \BibitemOpen
  \bibfield  {author} {\bibinfo {author} {\bibfnamefont {J.}~\bibnamefont
  {Sakstein}},\ }\href {\doibase 10.1103/PhysRevD.92.124045} {\bibfield
  {journal} {\bibinfo  {journal} {Phys. Rev.}\ }\textbf {\bibinfo {volume}
  {D92}},\ \bibinfo {pages} {124045} (\bibinfo {year} {2015}{\natexlab{c}})},\
  \Eprint {http://arxiv.org/abs/1511.01685} {arXiv:1511.01685 [astro-ph.CO]}
  \BibitemShut {NoStop}%
\bibitem [{\citenamefont {Sakstein}\ and\ \citenamefont
  {Koyama}(2015)}]{Sakstein:2015aqx}%
  \BibitemOpen
  \bibfield  {author} {\bibinfo {author} {\bibfnamefont {J.}~\bibnamefont
  {Sakstein}}\ and\ \bibinfo {author} {\bibfnamefont {K.}~\bibnamefont
  {Koyama}},\ }\href {\doibase 10.1142/S0218271815440216} {\bibfield  {journal}
  {\bibinfo  {journal} {Int. J. Mod. Phys.}\ }\textbf {\bibinfo {volume}
  {D24}},\ \bibinfo {pages} {1544021} (\bibinfo {year} {2015})}\BibitemShut
  {NoStop}%
\bibitem [{\citenamefont {Jain}\ \emph {et~al.}(2016)\citenamefont {Jain},
  \citenamefont {Kouvaris},\ and\ \citenamefont {Nielsen}}]{Jain:2015edg}%
  \BibitemOpen
  \bibfield  {author} {\bibinfo {author} {\bibfnamefont {R.~K.}\ \bibnamefont
  {Jain}}, \bibinfo {author} {\bibfnamefont {C.}~\bibnamefont {Kouvaris}}, \
  and\ \bibinfo {author} {\bibfnamefont {N.~G.}\ \bibnamefont {Nielsen}},\
  }\href {\doibase 10.1103/PhysRevLett.116.151103} {\bibfield  {journal}
  {\bibinfo  {journal} {Phys. Rev. Lett.}\ }\textbf {\bibinfo {volume} {116}},\
  \bibinfo {pages} {151103} (\bibinfo {year} {2016})},\ \Eprint
  {http://arxiv.org/abs/1512.05946} {arXiv:1512.05946 [astro-ph.CO]}
  \BibitemShut {NoStop}%
\bibitem [{\citenamefont {Sakstein}\ \emph {et~al.}(2016)\citenamefont
  {Sakstein}, \citenamefont {Wilcox}, \citenamefont {Bacon}, \citenamefont
  {Koyama},\ and\ \citenamefont {Nichol}}]{Sakstein:2016ggl}%
  \BibitemOpen
  \bibfield  {author} {\bibinfo {author} {\bibfnamefont {J.}~\bibnamefont
  {Sakstein}}, \bibinfo {author} {\bibfnamefont {H.}~\bibnamefont {Wilcox}},
  \bibinfo {author} {\bibfnamefont {D.}~\bibnamefont {Bacon}}, \bibinfo
  {author} {\bibfnamefont {K.}~\bibnamefont {Koyama}}, \ and\ \bibinfo {author}
  {\bibfnamefont {R.~C.}\ \bibnamefont {Nichol}},\ }\href {\doibase
  10.1088/1475-7516/2016/07/019} {\bibfield  {journal} {\bibinfo  {journal}
  {JCAP}\ }\textbf {\bibinfo {volume} {1607}},\ \bibinfo {pages} {019}
  (\bibinfo {year} {2016})},\ \Eprint {http://arxiv.org/abs/1603.06368}
  {arXiv:1603.06368 [astro-ph.CO]} \BibitemShut {NoStop}%
\bibitem [{\citenamefont {Babichev}\ \emph {et~al.}(2016)\citenamefont
  {Babichev}, \citenamefont {Koyama}, \citenamefont {Langlois}, \citenamefont
  {Saito},\ and\ \citenamefont {Sakstein}}]{Babichev:2016jom}%
  \BibitemOpen
  \bibfield  {author} {\bibinfo {author} {\bibfnamefont {E.}~\bibnamefont
  {Babichev}}, \bibinfo {author} {\bibfnamefont {K.}~\bibnamefont {Koyama}},
  \bibinfo {author} {\bibfnamefont {D.}~\bibnamefont {Langlois}}, \bibinfo
  {author} {\bibfnamefont {R.}~\bibnamefont {Saito}}, \ and\ \bibinfo {author}
  {\bibfnamefont {J.}~\bibnamefont {Sakstein}},\ }\href {\doibase
  10.1088/0264-9381/33/23/235014} {\bibfield  {journal} {\bibinfo  {journal}
  {Class. Quant. Grav.}\ }\textbf {\bibinfo {volume} {33}},\ \bibinfo {pages}
  {235014} (\bibinfo {year} {2016})},\ \Eprint
  {http://arxiv.org/abs/1606.06627} {arXiv:1606.06627 [gr-qc]} \BibitemShut
  {NoStop}%
\bibitem [{\citenamefont {Sakstein}\ \emph
  {et~al.}(2017{\natexlab{a}})\citenamefont {Sakstein}, \citenamefont
  {Babichev}, \citenamefont {Koyama}, \citenamefont {Langlois},\ and\
  \citenamefont {Saito}}]{Sakstein:2016oel}%
  \BibitemOpen
  \bibfield  {author} {\bibinfo {author} {\bibfnamefont {J.}~\bibnamefont
  {Sakstein}}, \bibinfo {author} {\bibfnamefont {E.}~\bibnamefont {Babichev}},
  \bibinfo {author} {\bibfnamefont {K.}~\bibnamefont {Koyama}}, \bibinfo
  {author} {\bibfnamefont {D.}~\bibnamefont {Langlois}}, \ and\ \bibinfo
  {author} {\bibfnamefont {R.}~\bibnamefont {Saito}},\ }\href {\doibase
  10.1103/PhysRevD.95.064013} {\bibfield  {journal} {\bibinfo  {journal} {Phys.
  Rev.}\ }\textbf {\bibinfo {volume} {D95}},\ \bibinfo {pages} {064013}
  (\bibinfo {year} {2017}{\natexlab{a}})},\ \Eprint
  {http://arxiv.org/abs/1612.04263} {arXiv:1612.04263 [gr-qc]} \BibitemShut
  {NoStop}%
\bibitem [{\citenamefont {Sakstein}\ \emph
  {et~al.}(2017{\natexlab{b}})\citenamefont {Sakstein}, \citenamefont
  {Kenna-Allison},\ and\ \citenamefont {Koyama}}]{Sakstein:2016lyj}%
  \BibitemOpen
  \bibfield  {author} {\bibinfo {author} {\bibfnamefont {J.}~\bibnamefont
  {Sakstein}}, \bibinfo {author} {\bibfnamefont {M.}~\bibnamefont
  {Kenna-Allison}}, \ and\ \bibinfo {author} {\bibfnamefont {K.}~\bibnamefont
  {Koyama}},\ }\href {\doibase 10.1088/1475-7516/2017/03/007} {\bibfield
  {journal} {\bibinfo  {journal} {JCAP}\ }\textbf {\bibinfo {volume} {1703}},\
  \bibinfo {pages} {007} (\bibinfo {year} {2017}{\natexlab{b}})},\ \Eprint
  {http://arxiv.org/abs/1611.01062} {arXiv:1611.01062 [gr-qc]} \BibitemShut
  {NoStop}%
\bibitem [{\citenamefont {Burrows}\ and\ \citenamefont
  {Liebert}(1993)}]{Burrows:1992fg}%
  \BibitemOpen
  \bibfield  {author} {\bibinfo {author} {\bibfnamefont {A.}~\bibnamefont
  {Burrows}}\ and\ \bibinfo {author} {\bibfnamefont {J.}~\bibnamefont
  {Liebert}},\ }\href {\doibase 10.1103/RevModPhys.65.301} {\bibfield
  {journal} {\bibinfo  {journal} {Rev. Mod. Phys.}\ }\textbf {\bibinfo {volume}
  {65}},\ \bibinfo {pages} {301} (\bibinfo {year} {1993})}\BibitemShut
  {NoStop}%
\bibitem [{\citenamefont {Hoekstra}\ \emph {et~al.}(2015)\citenamefont
  {Hoekstra}, \citenamefont {Herbonnet}, \citenamefont {Muzzin}, \citenamefont
  {Babul}, \citenamefont {Mahdavi}, \citenamefont {Viola},\ and\ \citenamefont
  {Cacciato}}]{Hoekstra:2015gda}%
  \BibitemOpen
  \bibfield  {author} {\bibinfo {author} {\bibfnamefont {H.}~\bibnamefont
  {Hoekstra}}, \bibinfo {author} {\bibfnamefont {R.}~\bibnamefont {Herbonnet}},
  \bibinfo {author} {\bibfnamefont {A.}~\bibnamefont {Muzzin}}, \bibinfo
  {author} {\bibfnamefont {A.}~\bibnamefont {Babul}}, \bibinfo {author}
  {\bibfnamefont {A.}~\bibnamefont {Mahdavi}}, \bibinfo {author} {\bibfnamefont
  {M.}~\bibnamefont {Viola}}, \ and\ \bibinfo {author} {\bibfnamefont
  {M.}~\bibnamefont {Cacciato}},\ }\href {\doibase 10.1093/mnras/stv275}
  {\bibfield  {journal} {\bibinfo  {journal} {Mon. Not. Roy. Astron. Soc.}\
  }\textbf {\bibinfo {volume} {449}},\ \bibinfo {pages} {685} (\bibinfo {year}
  {2015})},\ \Eprint {http://arxiv.org/abs/1502.01883} {arXiv:1502.01883
  [astro-ph.CO]} \BibitemShut {NoStop}%
\bibitem [{\citenamefont {Smith}\ \emph {et~al.}(2016)\citenamefont {Smith}
  \emph {et~al.}}]{Smith:2015qhs}%
  \BibitemOpen
  \bibfield  {author} {\bibinfo {author} {\bibfnamefont {G.~P.}\ \bibnamefont
  {Smith}} \emph {et~al.},\ }\href {\doibase 10.1093/mnrasl/slv175} {\bibfield
  {journal} {\bibinfo  {journal} {Mon. Not. Roy. Astron. Soc.}\ }\textbf
  {\bibinfo {volume} {456}},\ \bibinfo {pages} {L74} (\bibinfo {year}
  {2016})},\ \Eprint {http://arxiv.org/abs/1511.01919} {arXiv:1511.01919
  [astro-ph.CO]} \BibitemShut {NoStop}%
\bibitem [{\citenamefont {Applegate}\ \emph {et~al.}(2016)\citenamefont
  {Applegate} \emph {et~al.}}]{Applegate:2015kua}%
  \BibitemOpen
  \bibfield  {author} {\bibinfo {author} {\bibfnamefont {D.~E.}\ \bibnamefont
  {Applegate}} \emph {et~al.},\ }\href {\doibase 10.1093/mnras/stw005}
  {\bibfield  {journal} {\bibinfo  {journal} {Mon. Not. Roy. Astron. Soc.}\
  }\textbf {\bibinfo {volume} {457}},\ \bibinfo {pages} {1522} (\bibinfo {year}
  {2016})},\ \Eprint {http://arxiv.org/abs/1509.02162} {arXiv:1509.02162
  [astro-ph.CO]} \BibitemShut {NoStop}%
\bibitem [{\citenamefont {Zumalacárregui}\ and\ \citenamefont
  {García-Bellido}(2014)}]{Zumalacarregui:2013pma}%
  \BibitemOpen
  \bibfield  {author} {\bibinfo {author} {\bibfnamefont {M.}~\bibnamefont
  {Zumalacárregui}}\ and\ \bibinfo {author} {\bibfnamefont {J.}~\bibnamefont
  {García-Bellido}},\ }\href {\doibase 10.1103/PhysRevD.89.064046} {\bibfield
  {journal} {\bibinfo  {journal} {Phys. Rev.}\ }\textbf {\bibinfo {volume}
  {D89}},\ \bibinfo {pages} {064046} (\bibinfo {year} {2014})},\ \Eprint
  {http://arxiv.org/abs/1308.4685} {arXiv:1308.4685 [gr-qc]} \BibitemShut
  {NoStop}%
\bibitem [{\citenamefont {Deffayet}\ \emph {et~al.}(2015)\citenamefont
  {Deffayet}, \citenamefont {Esposito-Farese},\ and\ \citenamefont
  {Steer}}]{Deffayet:2015qwa}%
  \BibitemOpen
  \bibfield  {author} {\bibinfo {author} {\bibfnamefont {C.}~\bibnamefont
  {Deffayet}}, \bibinfo {author} {\bibfnamefont {G.}~\bibnamefont
  {Esposito-Farese}}, \ and\ \bibinfo {author} {\bibfnamefont {D.~A.}\
  \bibnamefont {Steer}},\ }\href {\doibase 10.1103/PhysRevD.92.084013}
  {\bibfield  {journal} {\bibinfo  {journal} {Phys. Rev.}\ }\textbf {\bibinfo
  {volume} {D92}},\ \bibinfo {pages} {084013} (\bibinfo {year} {2015})},\
  \Eprint {http://arxiv.org/abs/1506.01974} {arXiv:1506.01974 [gr-qc]}
  \BibitemShut {NoStop}%
\bibitem [{\citenamefont {Langlois}\ and\ \citenamefont
  {Noui}(2016)}]{Langlois:2015cwa}%
  \BibitemOpen
  \bibfield  {author} {\bibinfo {author} {\bibfnamefont {D.}~\bibnamefont
  {Langlois}}\ and\ \bibinfo {author} {\bibfnamefont {K.}~\bibnamefont
  {Noui}},\ }\href {\doibase 10.1088/1475-7516/2016/02/034} {\bibfield
  {journal} {\bibinfo  {journal} {JCAP}\ }\textbf {\bibinfo {volume} {1602}},\
  \bibinfo {pages} {034} (\bibinfo {year} {2016})},\ \Eprint
  {http://arxiv.org/abs/1510.06930} {arXiv:1510.06930 [gr-qc]} \BibitemShut
  {NoStop}%
\bibitem [{\citenamefont {Crisostomi}\ \emph {et~al.}(2016)\citenamefont
  {Crisostomi}, \citenamefont {Hull}, \citenamefont {Koyama},\ and\
  \citenamefont {Tasinato}}]{Crisostomi:2016tcp}%
  \BibitemOpen
  \bibfield  {author} {\bibinfo {author} {\bibfnamefont {M.}~\bibnamefont
  {Crisostomi}}, \bibinfo {author} {\bibfnamefont {M.}~\bibnamefont {Hull}},
  \bibinfo {author} {\bibfnamefont {K.}~\bibnamefont {Koyama}}, \ and\ \bibinfo
  {author} {\bibfnamefont {G.}~\bibnamefont {Tasinato}},\ }\href {\doibase
  10.1088/1475-7516/2016/03/038} {\bibfield  {journal} {\bibinfo  {journal}
  {JCAP}\ }\textbf {\bibinfo {volume} {1603}},\ \bibinfo {pages} {038}
  (\bibinfo {year} {2016})},\ \Eprint {http://arxiv.org/abs/1601.04658}
  {arXiv:1601.04658 [hep-th]} \BibitemShut {NoStop}%
\bibitem [{\citenamefont {Horndeski}(1974)}]{Horndeski:1974wa}%
  \BibitemOpen
  \bibfield  {author} {\bibinfo {author} {\bibfnamefont {G.~W.}\ \bibnamefont
  {Horndeski}},\ }\href {\doibase 10.1007/BF01807638} {\bibfield  {journal}
  {\bibinfo  {journal} {Int. J. Theor. Phys.}\ }\textbf {\bibinfo {volume}
  {10}},\ \bibinfo {pages} {363} (\bibinfo {year} {1974})}\BibitemShut
  {NoStop}%
\bibitem [{\citenamefont {Deffayet}\ \emph {et~al.}(2011)\citenamefont
  {Deffayet}, \citenamefont {Gao}, \citenamefont {Steer},\ and\ \citenamefont
  {Zahariade}}]{Deffayet:2011gz}%
  \BibitemOpen
  \bibfield  {author} {\bibinfo {author} {\bibfnamefont {C.}~\bibnamefont
  {Deffayet}}, \bibinfo {author} {\bibfnamefont {X.}~\bibnamefont {Gao}},
  \bibinfo {author} {\bibfnamefont {D.~A.}\ \bibnamefont {Steer}}, \ and\
  \bibinfo {author} {\bibfnamefont {G.}~\bibnamefont {Zahariade}},\ }\href
  {\doibase 10.1103/PhysRevD.84.064039} {\bibfield  {journal} {\bibinfo
  {journal} {Phys. Rev.}\ }\textbf {\bibinfo {volume} {D84}},\ \bibinfo {pages}
  {064039} (\bibinfo {year} {2011})},\ \Eprint {http://arxiv.org/abs/1103.3260}
  {arXiv:1103.3260 [hep-th]} \BibitemShut {NoStop}%
\bibitem [{\citenamefont {Deffayet}\ \emph {et~al.}(2009)\citenamefont
  {Deffayet}, \citenamefont {Esposito-Farese},\ and\ \citenamefont
  {Vikman}}]{Deffayet:2009wt}%
  \BibitemOpen
  \bibfield  {author} {\bibinfo {author} {\bibfnamefont {C.}~\bibnamefont
  {Deffayet}}, \bibinfo {author} {\bibfnamefont {G.}~\bibnamefont
  {Esposito-Farese}}, \ and\ \bibinfo {author} {\bibfnamefont {A.}~\bibnamefont
  {Vikman}},\ }\href {\doibase 10.1103/PhysRevD.79.084003} {\bibfield
  {journal} {\bibinfo  {journal} {Phys. Rev.}\ }\textbf {\bibinfo {volume}
  {D79}},\ \bibinfo {pages} {084003} (\bibinfo {year} {2009})},\ \Eprint
  {http://arxiv.org/abs/0901.1314} {arXiv:0901.1314 [hep-th]} \BibitemShut
  {NoStop}%
\bibitem [{\citenamefont {Brax}\ \emph {et~al.}(2016)\citenamefont {Brax},
  \citenamefont {Burrage},\ and\ \citenamefont {Davis}}]{Brax:2015dma}%
  \BibitemOpen
  \bibfield  {author} {\bibinfo {author} {\bibfnamefont {P.}~\bibnamefont
  {Brax}}, \bibinfo {author} {\bibfnamefont {C.}~\bibnamefont {Burrage}}, \
  and\ \bibinfo {author} {\bibfnamefont {A.-C.}\ \bibnamefont {Davis}},\ }\href
  {\doibase 10.1088/1475-7516/2016/03/004} {\bibfield  {journal} {\bibinfo
  {journal} {JCAP}\ }\textbf {\bibinfo {volume} {1603}},\ \bibinfo {pages}
  {004} (\bibinfo {year} {2016})},\ \Eprint {http://arxiv.org/abs/1510.03701}
  {arXiv:1510.03701 [gr-qc]} \BibitemShut {NoStop}%
\bibitem [{\citenamefont {Appleby}\ and\ \citenamefont
  {Linder}(2012)}]{Appleby:2011aa}%
  \BibitemOpen
  \bibfield  {author} {\bibinfo {author} {\bibfnamefont {S.}~\bibnamefont
  {Appleby}}\ and\ \bibinfo {author} {\bibfnamefont {E.~V.}\ \bibnamefont
  {Linder}},\ }\href {\doibase 10.1088/1475-7516/2012/03/043} {\bibfield
  {journal} {\bibinfo  {journal} {JCAP}\ }\textbf {\bibinfo {volume} {1203}},\
  \bibinfo {pages} {043} (\bibinfo {year} {2012})},\ \Eprint
  {http://arxiv.org/abs/1112.1981} {arXiv:1112.1981 [astro-ph.CO]} \BibitemShut
  {NoStop}%
\bibitem [{\citenamefont {Sakstein}\ \emph
  {et~al.}(2017{\natexlab{c}})\citenamefont {Sakstein}, \citenamefont {Jain},
  \citenamefont {Heyl},\ and\ \citenamefont {Hui}}]{Sakstein:2017bws}%
  \BibitemOpen
  \bibfield  {author} {\bibinfo {author} {\bibfnamefont {J.}~\bibnamefont
  {Sakstein}}, \bibinfo {author} {\bibfnamefont {B.}~\bibnamefont {Jain}},
  \bibinfo {author} {\bibfnamefont {J.~S.}\ \bibnamefont {Heyl}}, \ and\
  \bibinfo {author} {\bibfnamefont {L.}~\bibnamefont {Hui}},\ }\href {\doibase
  10.3847/2041-8213/aa7e26} {\bibfield  {journal} {\bibinfo  {journal}
  {Astrophys. J.}\ }\textbf {\bibinfo {volume} {844}},\ \bibinfo {pages} {L14}
  (\bibinfo {year} {2017}{\natexlab{c}})},\ \Eprint
  {http://arxiv.org/abs/1704.02425} {arXiv:1704.02425 [astro-ph.CO]}
  \BibitemShut {NoStop}%
\bibitem [{\citenamefont {Hui}\ and\ \citenamefont
  {Nicolis}(2013)}]{Hui:2012qt}%
  \BibitemOpen
  \bibfield  {author} {\bibinfo {author} {\bibfnamefont {L.}~\bibnamefont
  {Hui}}\ and\ \bibinfo {author} {\bibfnamefont {A.}~\bibnamefont {Nicolis}},\
  }\href {\doibase 10.1103/PhysRevLett.110.241104} {\bibfield  {journal}
  {\bibinfo  {journal} {Phys. Rev. Lett.}\ }\textbf {\bibinfo {volume} {110}},\
  \bibinfo {pages} {241104} (\bibinfo {year} {2013})},\ \Eprint
  {http://arxiv.org/abs/1202.1296} {arXiv:1202.1296 [hep-th]} \BibitemShut
  {NoStop}%
\bibitem [{\citenamefont {Hui}\ and\ \citenamefont
  {Nicolis}(2012)}]{Hui:2012jb}%
  \BibitemOpen
  \bibfield  {author} {\bibinfo {author} {\bibfnamefont {L.}~\bibnamefont
  {Hui}}\ and\ \bibinfo {author} {\bibfnamefont {A.}~\bibnamefont {Nicolis}},\
  }\href {\doibase 10.1103/PhysRevLett.109.051304} {\bibfield  {journal}
  {\bibinfo  {journal} {Phys. Rev. Lett.}\ }\textbf {\bibinfo {volume} {109}},\
  \bibinfo {pages} {051304} (\bibinfo {year} {2012})},\ \Eprint
  {http://arxiv.org/abs/1201.1508} {arXiv:1201.1508 [astro-ph.CO]} \BibitemShut
  {NoStop}%
\bibitem [{\citenamefont {Asvathaman}\ \emph {et~al.}(2017)\citenamefont
  {Asvathaman}, \citenamefont {Heyl},\ and\ \citenamefont
  {Hui}}]{Asvathaman:2015nna}%
  \BibitemOpen
  \bibfield  {author} {\bibinfo {author} {\bibfnamefont {A.}~\bibnamefont
  {Asvathaman}}, \bibinfo {author} {\bibfnamefont {J.~S.}\ \bibnamefont
  {Heyl}}, \ and\ \bibinfo {author} {\bibfnamefont {L.}~\bibnamefont {Hui}},\
  }\href {\doibase 10.1093/mnras/stw2905} {\bibfield  {journal} {\bibinfo
  {journal} {Mon. Not. Roy. Astron. Soc.}\ }\textbf {\bibinfo {volume} {465}},\
  \bibinfo {pages} {3261} (\bibinfo {year} {2017})},\ \Eprint
  {http://arxiv.org/abs/1506.07607} {arXiv:1506.07607 [astro-ph.GA]}
  \BibitemShut {NoStop}%
\bibitem [{\citenamefont {Kimura}\ \emph {et~al.}(2012)\citenamefont {Kimura},
  \citenamefont {Kobayashi},\ and\ \citenamefont {Yamamoto}}]{Kimura:2011dc}%
  \BibitemOpen
  \bibfield  {author} {\bibinfo {author} {\bibfnamefont {R.}~\bibnamefont
  {Kimura}}, \bibinfo {author} {\bibfnamefont {T.}~\bibnamefont {Kobayashi}}, \
  and\ \bibinfo {author} {\bibfnamefont {K.}~\bibnamefont {Yamamoto}},\ }\href
  {\doibase 10.1103/PhysRevD.85.024023} {\bibfield  {journal} {\bibinfo
  {journal} {Phys. Rev.}\ }\textbf {\bibinfo {volume} {D85}},\ \bibinfo {pages}
  {024023} (\bibinfo {year} {2012})},\ \Eprint {http://arxiv.org/abs/1111.6749}
  {arXiv:1111.6749 [astro-ph.CO]} \BibitemShut {NoStop}%
\bibitem [{\citenamefont {Koyama}\ \emph {et~al.}(2013)\citenamefont {Koyama},
  \citenamefont {Niz},\ and\ \citenamefont {Tasinato}}]{Koyama:2013paa}%
  \BibitemOpen
  \bibfield  {author} {\bibinfo {author} {\bibfnamefont {K.}~\bibnamefont
  {Koyama}}, \bibinfo {author} {\bibfnamefont {G.}~\bibnamefont {Niz}}, \ and\
  \bibinfo {author} {\bibfnamefont {G.}~\bibnamefont {Tasinato}},\ }\href
  {\doibase 10.1103/PhysRevD.88.021502} {\bibfield  {journal} {\bibinfo
  {journal} {Phys. Rev.}\ }\textbf {\bibinfo {volume} {D88}},\ \bibinfo {pages}
  {021502} (\bibinfo {year} {2013})},\ \Eprint {http://arxiv.org/abs/1305.0279}
  {arXiv:1305.0279 [hep-th]} \BibitemShut {NoStop}%
\bibitem [{\citenamefont {Kaloper}\ \emph {et~al.}(2015)\citenamefont
  {Kaloper}, \citenamefont {Padilla}, \citenamefont {Saffin},\ and\
  \citenamefont {Stefanyszyn}}]{Kaloper:2014vqa}%
  \BibitemOpen
  \bibfield  {author} {\bibinfo {author} {\bibfnamefont {N.}~\bibnamefont
  {Kaloper}}, \bibinfo {author} {\bibfnamefont {A.}~\bibnamefont {Padilla}},
  \bibinfo {author} {\bibfnamefont {P.}~\bibnamefont {Saffin}}, \ and\ \bibinfo
  {author} {\bibfnamefont {D.}~\bibnamefont {Stefanyszyn}},\ }\href {\doibase
  10.1103/PhysRevD.91.045017} {\bibfield  {journal} {\bibinfo  {journal} {Phys.
  Rev.}\ }\textbf {\bibinfo {volume} {D91}},\ \bibinfo {pages} {045017}
  (\bibinfo {year} {2015})},\ \Eprint {http://arxiv.org/abs/1409.3243}
  {arXiv:1409.3243 [hep-th]} \BibitemShut {NoStop}%
\bibitem [{\citenamefont {Keltner}\ and\ \citenamefont
  {Tolley}(2015)}]{Keltner:2015xda}%
  \BibitemOpen
  \bibfield  {author} {\bibinfo {author} {\bibfnamefont {L.}~\bibnamefont
  {Keltner}}\ and\ \bibinfo {author} {\bibfnamefont {A.~J.}\ \bibnamefont
  {Tolley}},\ }\href@noop {} {\  (\bibinfo {year} {2015})},\ \Eprint
  {http://arxiv.org/abs/1502.05706} {arXiv:1502.05706 [hep-th]} \BibitemShut
  {NoStop}%
\bibitem [{\citenamefont {de~Rham}\ \emph
  {et~al.}(2017{\natexlab{b}})\citenamefont {de~Rham}, \citenamefont
  {Melville}, \citenamefont {Tolley},\ and\ \citenamefont
  {Zhou}}]{deRham:2017avq}%
  \BibitemOpen
  \bibfield  {author} {\bibinfo {author} {\bibfnamefont {C.}~\bibnamefont
  {de~Rham}}, \bibinfo {author} {\bibfnamefont {S.}~\bibnamefont {Melville}},
  \bibinfo {author} {\bibfnamefont {A.~J.}\ \bibnamefont {Tolley}}, \ and\
  \bibinfo {author} {\bibfnamefont {S.-Y.}\ \bibnamefont {Zhou}},\ }\href@noop
  {} {\  (\bibinfo {year} {2017}{\natexlab{b}})},\ \Eprint
  {http://arxiv.org/abs/1702.06134} {arXiv:1702.06134 [hep-th]} \BibitemShut
  {NoStop}%
\bibitem [{\citenamefont {de~Rham}\ \emph
  {et~al.}(2017{\natexlab{c}})\citenamefont {de~Rham}, \citenamefont
  {Melville}, \citenamefont {Tolley},\ and\ \citenamefont
  {Zhou}}]{deRham:2017imi}%
  \BibitemOpen
  \bibfield  {author} {\bibinfo {author} {\bibfnamefont {C.}~\bibnamefont
  {de~Rham}}, \bibinfo {author} {\bibfnamefont {S.}~\bibnamefont {Melville}},
  \bibinfo {author} {\bibfnamefont {A.~J.}\ \bibnamefont {Tolley}}, \ and\
  \bibinfo {author} {\bibfnamefont {S.-Y.}\ \bibnamefont {Zhou}},\ }\href
  {\doibase 10.1007/JHEP09(2017)072} {\bibfield  {journal} {\bibinfo  {journal}
  {JHEP}\ }\textbf {\bibinfo {volume} {09}},\ \bibinfo {pages} {072} (\bibinfo
  {year} {2017}{\natexlab{c}})},\ \Eprint {http://arxiv.org/abs/1702.08577}
  {arXiv:1702.08577 [hep-th]} \BibitemShut {NoStop}%
\bibitem [{\citenamefont {Millington}\ \emph {et~al.}(2017)\citenamefont
  {Millington}, \citenamefont {Niedermann},\ and\ \citenamefont
  {Padilla}}]{Millington:2017sea}%
  \BibitemOpen
  \bibfield  {author} {\bibinfo {author} {\bibfnamefont {P.}~\bibnamefont
  {Millington}}, \bibinfo {author} {\bibfnamefont {F.}~\bibnamefont
  {Niedermann}}, \ and\ \bibinfo {author} {\bibfnamefont {A.}~\bibnamefont
  {Padilla}},\ }\href@noop {} {\  (\bibinfo {year} {2017})},\ \Eprint
  {http://arxiv.org/abs/1707.06931} {arXiv:1707.06931 [hep-th]} \BibitemShut
  {NoStop}%
\bibitem [{\citenamefont {Barreira}\ \emph {et~al.}(2012)\citenamefont
  {Barreira}, \citenamefont {Li}, \citenamefont {Baugh},\ and\ \citenamefont
  {Pascoli}}]{Barreira:2012kk}%
  \BibitemOpen
  \bibfield  {author} {\bibinfo {author} {\bibfnamefont {A.}~\bibnamefont
  {Barreira}}, \bibinfo {author} {\bibfnamefont {B.}~\bibnamefont {Li}},
  \bibinfo {author} {\bibfnamefont {C.~M.}\ \bibnamefont {Baugh}}, \ and\
  \bibinfo {author} {\bibfnamefont {S.}~\bibnamefont {Pascoli}},\ }\href
  {\doibase 10.1103/PhysRevD.86.124016} {\bibfield  {journal} {\bibinfo
  {journal} {Phys. Rev.}\ }\textbf {\bibinfo {volume} {D86}},\ \bibinfo {pages}
  {124016} (\bibinfo {year} {2012})},\ \Eprint {http://arxiv.org/abs/1208.0600}
  {arXiv:1208.0600 [astro-ph.CO]} \BibitemShut {NoStop}%
\bibitem [{\citenamefont {Renk}\ \emph {et~al.}(2017)\citenamefont {Renk},
  \citenamefont {Zumalacárregui}, \citenamefont {Montanari},\ and\
  \citenamefont {Barreira}}]{Renk:2017rzu}%
  \BibitemOpen
  \bibfield  {author} {\bibinfo {author} {\bibfnamefont {J.}~\bibnamefont
  {Renk}}, \bibinfo {author} {\bibfnamefont {M.}~\bibnamefont
  {Zumalacárregui}}, \bibinfo {author} {\bibfnamefont {F.}~\bibnamefont
  {Montanari}}, \ and\ \bibinfo {author} {\bibfnamefont {A.}~\bibnamefont
  {Barreira}},\ }\href {\doibase 10.1088/1475-7516/2017/10/020} {\bibfield
  {journal} {\bibinfo  {journal} {JCAP}\ }\textbf {\bibinfo {volume} {1710}},\
  \bibinfo {pages} {020} (\bibinfo {year} {2017})},\ \Eprint
  {http://arxiv.org/abs/1707.02263} {arXiv:1707.02263 [astro-ph.CO]}
  \BibitemShut {NoStop}%
\bibitem [{\citenamefont {Kase}\ and\ \citenamefont
  {Tsujikawa}(2014)}]{Kase:2014yya}%
  \BibitemOpen
  \bibfield  {author} {\bibinfo {author} {\bibfnamefont {R.}~\bibnamefont
  {Kase}}\ and\ \bibinfo {author} {\bibfnamefont {S.}~\bibnamefont
  {Tsujikawa}},\ }\href {\doibase 10.1103/PhysRevD.90.044073} {\bibfield
  {journal} {\bibinfo  {journal} {Phys. Rev.}\ }\textbf {\bibinfo {volume}
  {D90}},\ \bibinfo {pages} {044073} (\bibinfo {year} {2014})},\ \Eprint
  {http://arxiv.org/abs/1407.0794} {arXiv:1407.0794 [hep-th]} \BibitemShut
  {NoStop}%
\bibitem [{\citenamefont {Baker}\ \emph {et~al.}(2017)\citenamefont {Baker},
  \citenamefont {Bellini}, \citenamefont {Ferreira}, \citenamefont {Lagos},
  \citenamefont {Noller},\ and\ \citenamefont {Sawicki}}]{Baker:2017hug}%
  \BibitemOpen
  \bibfield  {author} {\bibinfo {author} {\bibfnamefont {T.}~\bibnamefont
  {Baker}}, \bibinfo {author} {\bibfnamefont {E.}~\bibnamefont {Bellini}},
  \bibinfo {author} {\bibfnamefont {P.~G.}\ \bibnamefont {Ferreira}}, \bibinfo
  {author} {\bibfnamefont {M.}~\bibnamefont {Lagos}}, \bibinfo {author}
  {\bibfnamefont {J.}~\bibnamefont {Noller}}, \ and\ \bibinfo {author}
  {\bibfnamefont {I.}~\bibnamefont {Sawicki}},\ }\href@noop {} {\  (\bibinfo
  {year} {2017})},\ \Eprint {http://arxiv.org/abs/1710.06394} {arXiv:1710.06394
  [astro-ph.CO]} \BibitemShut {NoStop}%
\bibitem [{\citenamefont {Ezquiaga}\ and\ \citenamefont
  {Zumalacárregui}(2017)}]{Ezquiaga:2017ekz}%
  \BibitemOpen
  \bibfield  {author} {\bibinfo {author} {\bibfnamefont {J.~M.}\ \bibnamefont
  {Ezquiaga}}\ and\ \bibinfo {author} {\bibfnamefont {M.}~\bibnamefont
  {Zumalacárregui}},\ }\href@noop {} {\  (\bibinfo {year} {2017})},\ \Eprint
  {http://arxiv.org/abs/1710.05901} {arXiv:1710.05901 [astro-ph.CO]}
  \BibitemShut {NoStop}%
\bibitem [{\citenamefont {Creminelli}\ and\ \citenamefont
  {Vernizzi}(2017)}]{Creminelli:2017sry}%
  \BibitemOpen
  \bibfield  {author} {\bibinfo {author} {\bibfnamefont {P.}~\bibnamefont
  {Creminelli}}\ and\ \bibinfo {author} {\bibfnamefont {F.}~\bibnamefont
  {Vernizzi}},\ }\href@noop {} {\  (\bibinfo {year} {2017})},\ \Eprint
  {http://arxiv.org/abs/1710.05877} {arXiv:1710.05877 [astro-ph.CO]}
  \BibitemShut {NoStop}%
\bibitem [{\citenamefont {Kanti}\ \emph {et~al.}(1996)\citenamefont {Kanti},
  \citenamefont {Mavromatos}, \citenamefont {Rizos}, \citenamefont {Tamvakis},\
  and\ \citenamefont {Winstanley}}]{Kanti:1995vq}%
  \BibitemOpen
  \bibfield  {author} {\bibinfo {author} {\bibfnamefont {P.}~\bibnamefont
  {Kanti}}, \bibinfo {author} {\bibfnamefont {N.~E.}\ \bibnamefont
  {Mavromatos}}, \bibinfo {author} {\bibfnamefont {J.}~\bibnamefont {Rizos}},
  \bibinfo {author} {\bibfnamefont {K.}~\bibnamefont {Tamvakis}}, \ and\
  \bibinfo {author} {\bibfnamefont {E.}~\bibnamefont {Winstanley}},\ }\href
  {\doibase 10.1103/PhysRevD.54.5049} {\bibfield  {journal} {\bibinfo
  {journal} {Phys. Rev.}\ }\textbf {\bibinfo {volume} {D54}},\ \bibinfo {pages}
  {5049} (\bibinfo {year} {1996})},\ \Eprint
  {http://arxiv.org/abs/hep-th/9511071} {arXiv:hep-th/9511071 [hep-th]}
  \BibitemShut {NoStop}%
\bibitem [{\citenamefont {van~de Bruck}\ \emph {et~al.}(2016)\citenamefont
  {van~de Bruck}, \citenamefont {Burrage},\ and\ \citenamefont
  {Morrice}}]{vandeBruck:2016cnh}%
  \BibitemOpen
  \bibfield  {author} {\bibinfo {author} {\bibfnamefont {C.}~\bibnamefont
  {van~de Bruck}}, \bibinfo {author} {\bibfnamefont {C.}~\bibnamefont
  {Burrage}}, \ and\ \bibinfo {author} {\bibfnamefont {J.}~\bibnamefont
  {Morrice}},\ }\href {\doibase 10.1088/1475-7516/2016/08/003} {\bibfield
  {journal} {\bibinfo  {journal} {JCAP}\ }\textbf {\bibinfo {volume} {1608}},\
  \bibinfo {pages} {003} (\bibinfo {year} {2016})},\ \Eprint
  {http://arxiv.org/abs/1605.03567} {arXiv:1605.03567 [gr-qc]} \BibitemShut
  {NoStop}%
\bibitem [{\citenamefont {Berezhiani}\ \emph {et~al.}(2017)\citenamefont
  {Berezhiani}, \citenamefont {Khoury},\ and\ \citenamefont
  {Wang}}]{Berezhiani:2016dne}%
  \BibitemOpen
  \bibfield  {author} {\bibinfo {author} {\bibfnamefont {L.}~\bibnamefont
  {Berezhiani}}, \bibinfo {author} {\bibfnamefont {J.}~\bibnamefont {Khoury}},
  \ and\ \bibinfo {author} {\bibfnamefont {J.}~\bibnamefont {Wang}},\ }\href
  {\doibase 10.1103/PhysRevD.95.123530} {\bibfield  {journal} {\bibinfo
  {journal} {Phys. Rev.}\ }\textbf {\bibinfo {volume} {D95}},\ \bibinfo {pages}
  {123530} (\bibinfo {year} {2017})},\ \Eprint
  {http://arxiv.org/abs/1612.00453} {arXiv:1612.00453 [hep-th]} \BibitemShut
  {NoStop}%
\bibitem [{\citenamefont {Copeland}\ \emph {et~al.}(2006)\citenamefont
  {Copeland}, \citenamefont {Sami},\ and\ \citenamefont
  {Tsujikawa}}]{Copeland:2006wr}%
  \BibitemOpen
  \bibfield  {author} {\bibinfo {author} {\bibfnamefont {E.~J.}\ \bibnamefont
  {Copeland}}, \bibinfo {author} {\bibfnamefont {M.}~\bibnamefont {Sami}}, \
  and\ \bibinfo {author} {\bibfnamefont {S.}~\bibnamefont {Tsujikawa}},\ }\href
  {\doibase 10.1142/S021827180600942X} {\bibfield  {journal} {\bibinfo
  {journal} {Int. J. Mod. Phys.}\ }\textbf {\bibinfo {volume} {D15}},\ \bibinfo
  {pages} {1753} (\bibinfo {year} {2006})},\ \Eprint
  {http://arxiv.org/abs/hep-th/0603057} {arXiv:hep-th/0603057 [hep-th]}
  \BibitemShut {NoStop}%
\bibitem [{\citenamefont {Brax}\ and\ \citenamefont
  {Burrage}(2014)}]{Brax:2014vva}%
  \BibitemOpen
  \bibfield  {author} {\bibinfo {author} {\bibfnamefont {P.}~\bibnamefont
  {Brax}}\ and\ \bibinfo {author} {\bibfnamefont {C.}~\bibnamefont {Burrage}},\
  }\href {\doibase 10.1103/PhysRevD.90.104009} {\bibfield  {journal} {\bibinfo
  {journal} {Phys. Rev.}\ }\textbf {\bibinfo {volume} {D90}},\ \bibinfo {pages}
  {104009} (\bibinfo {year} {2014})},\ \Eprint {http://arxiv.org/abs/1407.1861}
  {arXiv:1407.1861 [astro-ph.CO]} \BibitemShut {NoStop}%
\end{thebibliography}%
\bibliographystyle{apsrev4-1}

\end{document}